\documentclass[11pt,letterpaper]{article} 
\pdfoutput=1
\usepackage{amsmath,amssymb,amsfonts}
\usepackage[usenames,dvipsnames]{xcolor}
\definecolor{airforceblue}{rgb}{0.36, 0.54, 0.66}
\definecolor{meb}{rgb}{0.01, 0.31, 0.59}
\usepackage[T1]{fontenc}
\def\title{Quantum Null Geometry and Gravity} 
\usepackage{graphicx}
\usepackage[colorlinks=true, pdfstartview=FitV, linkcolor=blue, citecolor=magenta, urlcolor=blue]{hyperref}
\usepackage{cancel}
\usepackage{mdframed,framed}
\usepackage[usenames,dvipsnames]{xcolor}
\usepackage[normalem]{ulem}
\usepackage{layout}
\usepackage[nosort]{cite}

\newcommand{\zb}{\bar{z}}
\usepackage{textgreek}

\newcommand{\D}{\text{d}}
\newcommand{\beq}{\begin{eqnarray}}
\newcommand{\eeq}{\end{eqnarray}}
\newcommand{\beqn}{\begin{eqnarray}}
\newcommand{\eeqn}{\end{eqnarray}}
\newcommand{\bee}{\begin{equation} \begin{aligned}}
\newcommand{\eee}{ \end{aligned} \end{equation}}
\newcommand{\pa}{\partial}

\newcommand{\no}[1]{:\!{#1}\!:}
\newcommand{\RR}{\mathbb{R}}
\newcommand{\ZZ}{\mathbb{Z}}
\newcommand{\cL}{{\cal L}}
\newcommand{\cO}{{\cal O}}
\newcommand{\fL}{\mathfrak{L}}

\newcommand{\B}{\mathrm{B}}

\usepackage{tikz-cd}
\usepackage{pict2e}
\makeatletter
\newcommand{\variable@rule}[1]{%
  \fontdimen8  
  \ifx#1\displaystyle\textfont3\else
    \ifx#1\textstyle\textfont3\else
      \ifx#1\scriptstyle\scriptfont3\else
        \scriptscriptfont3\relax
  \fi\fi\fi
}
\usepackage{mathtools}

\usepackage{tocloft}
\newcommand{\ve}{\varepsilon}

\newcommand{\cN}{{\cal{N}}}
\newcommand{\cC}{{\cal{C}}}

\newcommand{\cF}{{\cal{F}}}
\newcommand{\cP}{{\cal{{P}}}}
\newcommand{\bcP}{{\overline{\cal{P}}}}
\newcommand{\bq}{{\overline{q}}}

\newcommand{\bX}{{\overline{X}}}
\newcommand{\bz}{{\overline{z}}}

\newcommand{\bzeta}{{\overline{\zeta}}}

\newcommand{\bD}{{\overline{\Delta}}}
\newcommand{\bN}{{\overline{N}}}
\newcommand{\rd}{\text{d}}

\newcommand{\emb}{\textepsilon\textmu\textbeta\textalpha \textdelta \'{o}\textnu}

\newcommand{\chkM}{{\color{red} \,\checkmark\kern-5pt{}_{M}}}

\newcommand{\be}{\begin{equation}}
\newcommand{\ee}{\end{equation}}
\newcommand{\bea}{\begin{eqnarray}}
\newcommand{\eea}{\end{eqnarray}}

\def\pa{\partial}

\newcommand{\dC}{ T }

\newcommand{\bC}{ {\pmb{C}} }
\newcommand{\bT}{ {\pmb{T}} }

\usepackage{ragged2e}
\usepackage{blindtext}

\newcommand{\perimeter}[1]{
	\centerline{
		\begin{minipage}[c]{0.7\textwidth}
			\begin{center}
			$^a$ Perimeter Institute for Theoretical Physics,\\
			 31 Caroline St. N., Waterloo ON, Canada, N2L 2Y5
			\end{center}
		\end{minipage}
		}
	}
\newcommand{\uiuc}[1]{
	\centerline{
		\begin{minipage}[c]{0.9\textwidth}
			\begin{center}
		$^b$ Illinois Center for Advanced Studies of the Universe \& Department of Physics,\\
			University of Illinois, 1110 West Green St., Urbana IL 61801, U.S.A.
			\end{center}
		\end{minipage}
		}
	}

\begin{document}

{\centering
 \vspace*{1cm}
\textbf{\LARGE{\title{}}}
\vspace{0.5cm}
\begin{center}
Luca Ciambelli,$^a$ Laurent Freidel,$^a$ Robert G. Leigh$^{b,a}$\\
\vspace{0.5cm}
\textit{\perimeter{}}\\
\vspace{0.5cm}
\textit{\uiuc{}}
\end{center}
\vspace{0.5cm}
{\small{\href{mailto:ciambelli.luca@gmail.com}{ciambelli.luca@gmail.com}, \ \href{mailto:lfreidel@perimeterinstitute.ca}{lfreidel@perimeterinstitute.ca}}}, \  \href{mailto:rgleigh@illinois.edu}{rgleigh@illinois.edu}
\vspace{2cm}
\begin{abstract}
\vspace{0.5cm}
In this work, we demonstrate that quantizing gravity on a null hypersurface leads to the emergence of a CFT associated with each null ray. This result stems from the ultralocal nature of null physics and is derived through a canonical analysis of the Raychaudhuri equation, interpreted as a constraint generating null time reparametrizations. The CFT exhibits a non-zero central charge, providing a mechanism for the quantum emergence of time in gravitational systems and an associated choice of vacuum state. Our analysis reveals that the central charge quantifies the degrees of freedom along each null ray. Throughout our investigation, the area element of a cut plays a crucial role, necessitating its treatment as a quantum operator due to its dynamic nature in phase space or because of quantum backreaction. Furthermore, we show that the total central charge diverges in a perturbative analysis due to the infinite number of null generators. This divergence is resolved if there is a discrete spectrum for the area form operator. We introduce the concept of `embadons' to denote these localized geometric units of area, the fundamental building blocks of geometry at a mesoscopic quantum gravity scale.
\end{abstract}}

\thispagestyle{empty}

\newpage
\tableofcontents
\thispagestyle{empty}
\newpage
\clearpage
\pagenumbering{arabic} 

\section{Introduction}

Physics on null hypersurfaces radically differs from physics on timelike and spacelike hypersurfaces. Already at the geometric level, a null hypersurface has a degenerate metric and a preferred vector field \cite{Henneaux:1979vn, Mars:1993mj, Gourgoulhon:2005ng}. The consequences of describing physical laws on this structure, now referred to as a Carrollian structure \cite{Duval:2014uva, Duval:2014uoa, Hartong:2015xda, Ciambelli:2019lap},\footnote{The Carroll group is the ultrarelativistic contraction of the Poincar\'e group \cite{LevyLeblond1965, Gupta1966}.} are still unfolding \cite{Penna:2018gfx, Ciambelli:2018wre, Freidel:2022bai, Freidel:2022vjq, Redondo-Yuste:2022czg, Petkou:2022bmz, deBoer:2023fnj, Adami:2023wbe}. Specifically, it took the community many decades to completely master the geometry and dynamics induced by gravity on a generic finite distance null hypersurface \cite{Damour:1979wya, Price:1986yy, Parikh:1997ma, Chandrasekaran:2018aop, Donnay:2019jiz, Adami:2020amw, Ciambelli:2023mir}. In \cite{Ciambelli:2023mir} (see as well  \cite{Chandrasekaran:2023vzb, Odak:2023pga}), echoing \cite{Torre:1985rw, Parattu:2015gga, Lehner:2016vdi, Hopfmuller:2016scf, DePaoli:2017sar, Hopfmuller:2018fni, Adami:2020ugu, Chandrasekaran:2021hxc, Adami:2021nnf, Sheikh-Jabbari:2022mqi}, a symplectic analysis for the Raychaudhuri constraint has been performed, and the Poisson bracket has been derived. 

While at asymptotic infinity the symplectic phase space has been quantized \cite{Ashtekar1981, Dappiaggi:2005ci, Ashtekar:2014zsa, Strominger:2017zoo, Ashtekar:2018lor, Hollands:2019ajl, Laddha:2020kvp, Prabhu:2022zcr, Prabhu:2024zwl, Prabhu:2024lmg}, at finite distance this task is more challenging. The quantization of finite-distance null hypersurfaces has been primarily addressed for black holes \cite{kay1991theorems, Wall:2011hj} and its thermodynamics \cite{Bekenstein:1972tm, Bekenstein:1973ur, Hawking:1975vcx,  Unruh:1976db, Bombelli:1986rw, Susskind:1994sm, Jacobson:2003wv, Solodukhin:2011gn},   while generic null hypersurfaces have been studied in the series of precursory works by Reisenberger \cite{Reisenberger:2007pq, Reisenberger:2007ku, Reisenberger:2012zq, Fuchs:2017jyk, Reisenberger:2018xkn}, Wieland \cite{Wieland:2017cmf, Wieland:2017zkf, Wieland:2019hkz, Wieland:2021vef, Wieland:2024dop}, and in  \cite{Speziale:2013ifa}. At finite distance, an important breakthrough led by Bousso and Wall has been the application of quantum information tools to QFT and gravity, leading to the formulation of entropy bounds and area laws \cite{Wall:2011hj, Wald:1991xn, Bousso:1999cb, Bousso:1999xy,  Bianchi:2012ev, Lewkowycz:2013nqa, Bousso:2014uxa, Bousso:2014sda, Bousso:2015mna, Bousso:2015wca, Faulkner:2016mzt, Casini:2017roe, Balakrishnan:2017bjg, Rignon-Bret:2023fjq, Hollands:2024vbe, Visser:2024pwz}. A major difference between finite distance and asymptotic infinity is that in the latter the Ashtekar-Streubel phase space (and generalizations thereof \cite{Barnich:2009se, Campiglia:2014yka, Compere:2018ylh, Campiglia:2020qvc, Freidel:2021fxf, Geiller:2022vto}) does not contain the celestial sphere metric, which is therefore not dynamical. This implies that one can successfully quantize asymptotic gravitons without quantizing the underlying geometry, even for finite Newton's constant. At finite distance the situation changes: the metric of a cut and the shear are conjugate pairs, and thus one cannot avoid promoting the geometric data to quantum operators. Even perturbatively in Newton's constant, if one wants to include quantum backreaction \cite{Larsen:1995ax, Burgess:2003jk, Donoghue:2012zc, Giddings:2015lla, Giddings:2022hba} (see also \cite{Thiemann:2024uwr} and references therein), the geometry must be properly quantized. These quantum effects predict that the  area is an operator \cite{Rovelli:1994ge,Ashtekar:1996eg} which needs to be identified with the modular Hamiltonian, and therefore acquires quantum fluctuations \cite{Bousso:2014uxa, Verlinde:2019xfb, Verlinde:2022hhs}.\footnote{Other signatures of quantum gravity effects on null hypersurfaces are described in \cite{Parikh:2020fhy, Mehdi:2022oct, Bak:2023wwo}.} The quantum aspects of null hypersurfaces and gravity-induced dynamics is the primary target of this manuscript. 

Although seemingly more technical, we focus our attention on the null case because some universal features drastically simplify the study of dynamics, such as the notion of ultralocality, which has been fully appreciated and exploited in \cite{Wall:2011hj}. It dictates that the Poisson brackets are zero for fields evaluated at different null rays, rendering the system effectively one-dimensional. In other words, fields on different null generators are independent. This is a tremendous simplification with dramatic consequences for the quantization. Another simplification is that null physics is inherently conformal, as there is no intrinsic meaning of distances in null time; the only meaningful notion is that of past or future, i.e., of the causality of events. This fact has been successfully exploited in near horizon physics, to appreciate that gravity contains a dimensionally-reducible CFT sector \cite{Solodukhin:1998tc, Carlip:2017eud, Gukov:2022oed}, which could provide a framework for computing the black hole entropy from horizon's microstates \cite{Strominger:1997eq, Carlip:1998wz, Solodukhin:1998tc, Carlip:1999cy}. However, this has been done only for specific backgrounds, such as spherically symmetric spacetimes and/or (near) extremal black holes, in which an AdS subspace emerges and thus a CFT naturally appears from its isometries (see e.g. \cite{Balasubramanian:2009bg}). It is our intention to show that this CFT structure pertains to all geometries: It is a universal feature of the induced gravitational phase space on a generic null hypersurface. Indeed, we will demonstrate that ultra-locality and conformality implies that every gravitational system is a one-dimensional (i.e., a two-dimensional chiral) CFT, once projected on a null hypersurface. This provides us with a window toward the quantization of null geometry and gravity, in which the infinitesimal area of a cut is pivotal. We emphasize that the dynamical fields are the geometric data themselves, together with the spin-$2$ gravitons and matter.

The most basic quantum feature of a two-dimensional chiral CFT is its central charge. We show that in this context, the central charge is associated with the boost symmetry, that is, the rescaling of the null vector field, and we compute its value explicitly in certain settings. Turning off the spin-$2$ radiative fields, we compute its contribution from the geometric data alone, which organizes as a $\beta\gamma$ CFT \cite{Losev:2005pu, Nekrasov:2005wg}. We can also compute the central charge in perturbative gravity, and we propose a novel framework here for considering perturbations around expanding backgrounds. We show that the central charge is a count of the number of fields on each null ray, and thus it diverges as long as there are infinitely many null generators on a cut. 

The central extension of the algebra of constraints is a key result. We have that the central charge vanishes classically, diverges in QFT, and we postulate that it is finite in quantum gravity. Furthermore, we argue that the central charge plays an important role in addressing the problem of time, which is one of the main issues of background-independent models of quantum gravity \cite{isham1993canonical, anderson2012problem, Carlip:2023daf}. Indeed, diffeomorphism invariance and background independence lead to the absence of a physical time, expressed in the impossibility of defining creation and annihilation operators. We take the appearance of a central charge and the subsequent anomaly in rescaling time as an indication that {\it time appears as a result of a quantum effect} . This is related to the constraint algebra being represented projectively and the appearance of a preferred $\mathfrak{sl}(2,\RR)$-invariant vacuum state.

The central charge being infinite is a manifestation of the universal nature of divergences in QFT \cite{bjorken1965relativistic}.
This is also present in any effective field theory of gravity, and it is rooted in the problem of subregions and factorization of the gravity phase space at the classical level, and Hilbert space at the quantum level, \cite{Giddings:2015lla, Connes:1994hv, Donnelly:2016rvo, Donnelly:2016auv, Witten:2018zxz, Witten:2021unn, Chandrasekaran:2022cip, Witten:2023qsv, Jensen:2023yxy, Klinger:2023auu, Gesteau:2023hbq, Kudler-Flam:2023qfl, Ball:2024hqe, Kudler-Flam:2024psh}.\footnote{This is particularly relevant in lower-dimensional gravity models, where one can fully quantize the theory. Recent papers include  \cite{Almheiri:2014cka, Maldacena:2016upp, Iliesiu:2019xuh, Mertens:2022irh, Kolchmeyer:2023gwa, Penington:2023dql, Iliesiu:2024cnh}.} A potential resolution is the introduction of quantum constituents that localize the degrees of freedom to subregions. The constituents are quantum bits of area, crucial in the quantization.
Indeed, the area universally couples the spin-$0$ geometric data to the spin-$2$ and matter, and its value classifies the different types of representation of the corner group associated with any codimension-$2$ cut of spacetime, \cite{Donnelly:2020xgu, Donnelly:2022kfs}. We therefore propose a molecular quantization in which the quantum area operator can vanish except at `punctures.' On the latter, the area form creates geometry, and thus, we refer to these area constituents as \emb, which is the Greek term for area and in the Latin alphabet it reads \emb = embadon. The works of Krasnov can be considered as progenitors of the embadon \cite{Smolin:1995vq,Krasnov:1996tb,Krasnov:1996wc}. This is similar to the introduction of molecules in the discovery of Brownian motion. If only $N$ punctures are created, we show that the total central charge becomes finite, such that the QFT limit is a large-$N$ limit.

The paper is organized as follows. In Section \ref{sec2} we review  \cite{Ciambelli:2023mir}, focusing on useful results for the quantum aspects of the theory. We then offer a classical application, the area commutator at two different null times. Finally, we  discuss the notion of conformal time and caustic avoiding times. Section \ref{sec:CFT} is devoted to the principle quantum computations of the manuscript. After proposing the OPE of the Raychaudhuri constraint at the quantum level, we focus on the spin-$0$ sector alone and find that its quantization leads to a $\beta\gamma$ CFT. Ultralocality then implies that the central charge diverges due to the infinite number of null generators. Afterwards, we reintroduce the spin-$2$ degrees of freedom in the perturbative regime $G\to 0$. We describe how to set up a phase space analysis for generic expanding backgrounds. Performing the canonical quantization of the system, we compute the central charge for non-expanding backgrounds and discuss the steps for expanding backgrounds. The section ends with the application of our results at asymptotic null infinity, where the infinite central charge is at the core of the infinite Bondi mass quantum fluctuations. In Section \ref{qutime}, we argue that the appearance of time is a quantum effect in gravity. This is rooted in the construction of gauge-invariant observables and in the choice of vacuum. In particular, the presence of an anomaly in the quantum constraint algebra is crucial for the appearance of time, and we argue that certain features of QFT are recovered if the central charge diverges. The central charge can be rendered finite by introducing the concept of embadon. In Section \ref{mol}, we show how we can achieve this while maintaining covariance. For this, we argue that the area at an intermediate energy scale, called mesoscopic gravity, has discrete support on the cuts, leading to a molecular view of the geometric data. We finish by comparing this approach with other quantum gravity scenarios. Section \ref{finale} offers a series of open questions and avenues of investigations that we intend to pursue next. Technical details on the derivation of the Poisson brackets are displayed in Appendixes \ref{class} and \ref{A2}.

\section{Classical Results}\label{sec2}

\subsection{Review of  Previous Results}

Let us begin with a short review of  \cite{Ciambelli:2023mir}. We will focus on concepts that are directly relevant to the present paper, and further details are collected in Appendix \ref{class}. 
Consider a $3$-dimensional manifold $\cN$ endowed with a Carrollian structure.  The geometric data are a nowhere-vanishing Carrollian vector field $\ell^a$ and a degenerate (corank-1) metric $q_{ab}$ such that $\ell^a q_{ab}=0$. This is the geometric baseline to describe a generic null hypersurface in spacetime from an intrinsic perspective, where $\ell^a$ describes the null generators. 

In \cite{Ciambelli:2023mir}, building upon previous results \cite{Ciambelli:2018wre, Ciambelli:2019lap, Chandrasekaran:2018aop, Donnay:2019jiz, Freidel:2022vjq}, 
we have seen that the bulk Einstein equations projected to a null hypersurface, i.e., the Raychaudhuri and Damour constraints, can be recast as the conservation of a  Carrollian energy-momentum tensor. That is, introducing the Carrollian connection $D_a$, the constraints can be written as $D_b T_a{}^b= T^{\mathrm{mat}}_{a\ell}$. If we project this on $\ell=\partial_v$, we obtain  in particular the Raychaudhuri equation
\beq\label{RC1}
C= T^{\mathrm{mat}}_{\ell \ell} -\ell^a D_b T_a{}^b =\pa_v^2\Omega-\mu\pa_v\Omega+\Omega(\sigma_{a}{}^{b}\sigma_{b}{}^a+8\pi G T^{\mathrm{mat}}_{vv}).
\eeq
Here, $\sigma_a{}^b$ is the shear, $\Omega$ is the infinitesimal area element on a cut $\cC$ of $\cN$, and $\mu$ is the \emph{surface tension}, defined as
\beq
\mu=\kappa+\frac{\theta}{2}.
\eeq 
It thus contains the inaffinity $\kappa$,\footnote{On $\cN$ we have $D_\ell \ell^a =\kappa \ell^a$.} which coincides with  the surface gravity in the special case of a  Killing horizon, and the expansion $\theta$. The surface tension, identified in \cite{Hopfmuller:2016scf}, plays a pivotal role in our analysis.

The manifold $\cN$ is a null hypersurface in the bulk. A generic null hypersurface is expected to develop caustics at finite time.\footnote{We generically refer to creases, corners, and caustics, as discussed in \cite{Gadioux:2023pmw}, as caustics.} Caustics are focusing points, where the null rays converge, and thus the area of transverse cuts shrinks to zero, which implies locally $\Omega \to 0$. We will choose $\cN$ to be the portion of a null hypersurface between two caustics, or between one caustic and infinity, such that classically $\Omega$ and thus its integral (the area of a cut) is always positive definite inside $\cN$. Calling the caustics $\cC_0$ and ${\cC}_1$, we therefore require $\partial {\cN}= {\cC}_0 \cup {\cC}_1$. As we will see, we can always perform a diffeomorphism to a  time that spans from $-\infty$ to $+\infty$ inside $\cN$. Thus, without loss of generality, we can assume that we are using coordinates such that $\cN$ is of infinite coordinate extent. One should note that given such a choice of time, there are diffeomorphisms that preserve these properties and there are diffeomorphisms that do not. We will see later that these remarks are important for quantization, and in fact one must make a judicious choice of time to be used. 

In \cite{Ciambelli:2023mir}, we constructed the primed canonical phase space, in which the condition $\delta \ell=0$ is imposed. As we review in Section \ref{prim}, this can be achieved by suitably combining the symmetries of the theory -- diffeomorphisms on $\cN$ and local internal boosts (rescalings of $\ell$) --,  leading to the primed phase space transformations
\beq\label{trs}
\fL'_{\hat{f}}q_{ab}=\cL_f q_{ab},\qquad
\fL'_{\hat{f}} \sigma_a{}^b =   \pa_v( f\sigma_a{}^b),\qquad\qquad\\
\fL'_{\hat{f}}\mu=\pa_v(f\mu )+\pa_v^2f\label{mutr},\qquad
\fL'_{\hat{f}}\Omega=f\pa_v\Omega\label{omtr},\qquad
\fL'_{\hat{f}}\theta=\pa_v(f\theta)\label{lft}
. \label{sitr}
\eeq
An important point is that the transformation of $\mu$ is anomalous, in the sense of \cite{Hopfmuller:2018fni}. That is, one might have expected $\mu$ to transform as a 1-form $\cL_{f} \mu=\pa_v(f\mu)$, but under the primed phase space transformation, it transforms as a  connection. One can say that its phase space transformation does not agree with the naive spacetime transformation $\Delta_f \mu = (\fL_{\hat{f}}-\cL_{f}) \mu = \pa_v^2 f$, where $\cL_{f}$ is the spacetime Lie derivative associated to $f\pa_v$, see \cite{ Freidel:2021cjp}. 

The canonical symplectic two form for the gravitational degrees of freedom is
\beq\label{ocan2}
\Omega^{\mathsf{can}} =\frac{1}{8\pi G}\int_{\cN}\ve_{\cN}^{(0)} \Big(\delta\left(\tfrac{1}{2}\Omega \sigma^{ab}\right)\wedge\delta q_{ab}-\delta\mu \wedge \delta\Omega\Big),
\eeq
where, since we are focusing on the Raychaudhuri constraint, we decoupled the spin-1 sector setting $\delta \ell=0$ (the primed phase space) and thus gauge fixed $\ell=\pa_v$. We introduced in this expression the coordinate measure $\ve_{\cN}^{(0)}=\rd v \wedge \ve_{\cC}^{(0)}$, where $\ve_{\cC}=\Omega\ve_{\cC}^{(0)}$ is the measure on a cut $\cC$ at constant $v$.

Using the field variations spelled out above one can then compute the charges
\beq
I'_{\hat f}\Omega^{\mathsf{can}}=-\delta M_f+\cF_f.
\eeq
The last term is a corner term that represents the symplectic flux
\beq\label{sympf}
\cF_{f}
=\frac1{8\pi G}\int_{{\partial\cN}}\ve_{\cC}^{(0)}f\left(\tfrac{1}{2}\Omega\sigma^{ab}\delta q_{ab}-\mu\delta\Omega\right),
\eeq
and needs to vanish in order for the symmetry to be canonical.\footnote{In this analysis we restrict our attention to cuts at constant $v$, since we have gauge fixed $\ell=\pa_v$. These cuts are also called corners in this manuscript, as they are codimension-$2$ surfaces from the bulk perspective.} The first term is the Noether charge
\beq\label{primech}
M_f = \frac1{8\pi G}\int_{\cN}  \ve^{(0)}_{\cN} \big(f C+\pa_v\left(\Omega \pa_v f-f\pa_v \Omega\right)\big)\;\hat =\; \frac1{8\pi G}\int_{{\partial\cN}}\ve^{(0)}_{\cC}\left(\Omega \pa_v f-f\pa_v \Omega\right).
\eeq
Here $\hat =$ means that we  impose the constraint $C=0$. In the absence of fluxes at the boundaries, we computed  the charge algebra
\beq
\{M_f,M_g\}=-M_{[f,g]},
\eeq
where for convenience we write the Witt Lie bracket as 
\beq\label{deffnbrac}
[f,g] :=f\pa_v g-g\pa_vf.
\eeq
This is the Lie algebra of  infinitesimal null time reparametrizations labelled  by functions $f$,  which represents the  vector field $f\pa_v$. 

We recall in Appendix \ref{class} the derivation of the Poisson bracket of elementary fields. Using these, we can evaluate the algebra of composite operators. By construction, the Raychaudhuri constraint satisfies the algebra
\beq\label{Calg}
\{ C_f,C_g\}=-8\pi G \ C_{[f,g]},
\eeq
where we introduced the current
\beq\label{Ccur}
C_f=\int_{\cN}\ve^{(0)}_{\cN} f C.
\eeq
These currents, with $C$ given by \eqref{RC1}, generate the transformations (\ref{trs}-\ref{sitr}), i.e. $\{C_f, O\} = 8\pi G\,  \fL'_{\hat{f}}O $.

Another crucial property of null hypersurfaces, and Carrollian physics in general, that can be deduced from these Poisson brackets is {\it ultralocality} on the cut. Indeed, the propagator \eqref{propa} displays a delta function on the cut $\delta^{(2)}(z_1-z_2)$: the bracket among fields on different null generators identically vanishes. This is an extremely powerful feature of  null physics; it implies that physics on a null hypersurface can always be reduced to a $1$-dimensional system, per each null generator. This ultralocal nature has been  appreciated in \cite{Wall:2011hj} and exploited to prove the generalized second law for black hole horizons. This property will be systematically exploited in the following, in making all results valid per null generator, and thus local on $\cC$. 

Classically the proper thing to do is to set the constraint to zero, as we did for example in \eqref{primech}. As we mentioned in \cite{Ciambelli:2023mir}, this has the appealing interpretation of setting a 1d (or more precisely a chiral 2d) stress tensor to zero. This stress tensor can be decomposed into constituent  spin-0, spin-2, and matter parts. The spin-$2$ sector contains the shear and thus the radiative degrees of freedom, while the spin-$0$ sector is given by $\Omega$ and $\mu$, and thus describes   geometric data. The Raychaudhuri constraint is then a balance equation between the spin-0 degrees of freedom and the radiative and matter degrees of freedom.

While this is an appealing picture, we will argue in this paper that the quantum version is quite different: the constraint algebra given above is centrally extended, there being a chiral central charge that would, as is familiar, appear in a number of places, particularly in the operator product expansion of currents. Because we wish to consider the theory coupled to arbitrary (massless) matter, this is not a situation where we would expect the total central charge to vanish. We will see that this central charge is crucial in the construction of a sensible quantum theory --- in fact, it is an important feature of the quantum theory rather than a sickness.

\subsection{Area Commutator}\label{sec:areacomms}

As an application of our established phase space structure we compute the Poisson bracket of the area at two different times. Thanks to ultralocality, we can perform this analysis for the infinitesimal area element, that is, per null generator. To compute this quantity, we introduce the following notation. We introduce the charge aspect $q_f$
\beq\label{chargasp}
M_f= \frac1{8\pi G}\int_{\cN}  \ve^{(0)}_{\cN} \big(f C+\pa_v\left(\Omega \pa_v f-f\pa_v \Omega\right)\big)\;\hat{=}\;\int_{\partial \cN}\ve_{\cC}^{(0)}q_f \qquad q_f:=\frac1{8\pi G}(\Omega \pa_v f-f\pa_v \Omega).
\eeq
We can evaluate the local charge between two null times $v_0$ and $v_1$, and we define the light-ray operator
\beq\label{lrop}
m_{v_0}^{v_1}[f](z,\bz) =\frac{1}{8\pi G}
\int_{v_0}^{v_1}\rd v \ \big(f C+\pa_v\left(\Omega \pa_v f-f\pa_v \Omega\right)\big)(x),
\eeq
where $x=(v,z,\bz)$ and $(z,\bz)$ denotes a point on the spatial cut. The latter dependence will be implied from now on.
On-shell, the Raychaudhuri constraint $C$ vanishes, and we are left with a corner charge
\beq
m_{v_0}^{v_1}[f]\;\hat{=}\;
\frac{1}{8\pi G}(\Omega\pa_vf-f\pa_v\Omega)\Big|^{v_1}_{v_0}=q_f(v_1)-q_f(v_0).
\eeq
Given the light-ray operator, we can extract $q_f(v_1)$ by arranging for $q_f(v_0)$ to vanish. This can be done in two different ways, either by requiring the vector field and its first derivative to vanish at $v_0$, or by requiring the cut at $v_0$ to have some particular properties. For instance, if there is a caustic at $v_0$, then $\Omega(v_0)=0$, and it is sufficient to require that the vector field vanishes there, $f(v_0)=0$, to obtain $q_f(v_0)=0$. Alternatively, suppose that the expansion vanishes at $v_0$ while the area remains finite, then $\pa_v \Omega (v_0)=0$, and thus it is sufficient to require that $\pa_v f(v_0)=0$. This is an interesting setup when discussing a black hole event horizon.

In the following, we will assume that a caustic is formed at $v_0$, and choose a vector field $f$ that vanishes there, such that we obtain 
\beq
m_{v_0}^{v_1} [f]\;\hat{=}\;
q_f(v_1).
\eeq
Importantly, we also suppose that the vector field is such that it has support only on the domain $[v_0,v_1]$, namely $f(v\geq v_1)=0$. If, in addition, the vector field $f$ is a boost for $v$ approaching $v_1$, normalized such that $\pa_vf(v_1)=1$, then 
\beq
m_{v_0}^{v_1}[f]\;\hat{=}\;\frac{1}{8\pi G}\Omega(v_1)
\eeq 
is the area density of the cut at $v_1$. A simple example of such a vector field is where $f(v)=\frac{v-v_0}{v_{10}}(v-v_1)$ on the domain $v\in [v_0,v_1]$, where we use the  notation $v_{ij}=v_i-v_j$. Near the corner $v\sim v_0$, this resembles a boost.\footnote{If, on the other hand, the expansion vanishes at $v_0$, we would require $\pa_vf(v_0)=0$ so that the charge aspect vanishes there, and we then obtain  
$m_{v_0}^{v_1}[f]\;\hat{=}\;\frac{\Omega(v_1)}{8\pi G}$.
An example of such a vector field is given by 
$f(v)=\frac{(v-v_0)^2}{2v_{10}}-\frac12 v_{10}$.
}
A basic observable is the Poisson bracket of two area densities at different null times. 
We can write the bracket of two light-ray operators in the form 
\beq\label{bracketvar}
\left\{ m_{v_0}^{v_1}[f_1],m_{v_0}^{v_2}[f_2]\right\}
= \Theta(v_{21}) \fL'_{\hat f_2} m^{v_1}_{v_0}[f_1] -
\Theta(v_{12})\fL'_{\hat f_1} m_{v_0}^{v_2}[f_2].
\eeq
This bracket implements a time ordering, where the light-ray operator defined on the larger interval acts in the phase space on the light-ray operator defined on the smaller interval. 
We notice that in the case where $v_1=v_2$, this reduces to the canonical equal-null-time bracket.
We thus can evaluate the Poisson bracket by considering each term separately. Let us assume $v_2\geq v_1$, and thus focus on the first term. The light-ray operator $m_{v_0}^{v_2}[f_2]$ is well-defined on the domain $[v_0, v_2]$. Since $f_1$ has support on $[v_0,v_1]\subset [v_0, v_2]$, it can be canonically  extended into a vector field $\tilde{f}_1$  on $[v_0, v_2]$ which is identical to $f_1$ on $[v_0,v_1]$  and  vanishes outside of it. Since $f_1(v_1)=0$ we have that the vector field $f_2$ involved in the variation acts smoothly on $\tilde{f}_1$. So we compute
\beqn
\fL'_{\hat f_2} m_{v_0}^{v_1}[f_1]
=\frac{1}{8\pi G}
\int_{v_0}^{v_1}\rd v
\Big[f_1\fL'_{\hat f_2} C+\pa_v(\fL'_{\hat f_2}\Omega\pa_v f_1-f_1\pa_v\fL'_{\hat f_2}\Omega)\Big].
\eeqn
Here, we implicitly used ultralocality and thus restricted our attention to a single null generator. These expressions should, however, be understood as local on the cut, with always a delta function $\delta^{(2)}(z_{12})=\delta^{(2)}(z_1-z_2)$ enforcing ultralocality.

We then evaluate
\beq
\fL'_{\hat f_2}C
=f_2\pa_vC+2\pa_vf_2C,\qquad
\fL'_{\hat f_2}\Omega=f_2\pa_v\Omega
\eeq
and thus we find
\beqn
\fL'_{\hat f_2} m_{v_0}^{v_1}[f_1]
&=&\frac{1}{8\pi G} \int_{v_0}^{v_1}\rd v
\Big[f_1f_2\pa_vC+2f_1\pa_vf_2C
+\pa_v(f_2\pa_vf_1\pa_v\Omega-f_1\pa_v(f_2\pa_v\Omega))\Big]\nonumber\\
&=&\frac{1}{8\pi G}
\int_{v_0}^{v_1}\rd v
\Big[\pa_v(f_1f_2C)+\{f_1,f_2\}C
-\pa_v(f_1f_2\pa_v^2\Omega+[f_1,f_2]\pa_v\Omega )\Big]\nonumber\\
&=&
m_{v_0}^{v_1}[[f_1,f_2]]+
\frac{1}{8\pi G}\Big[f_1f_2(C-\pa_v^2\Omega)-\Omega\pa_v[f_1,f_2]
\Big]_{v_0}^{v_1} \label{deltaarea}.
\eeqn
Alternatively, one can directly compute the Poisson bracket of areas using the kinematic Poisson brackets reviewed in Appendix \ref{class}, and one finds consistency with the result \eqref{deltaarea}, together with \eqref{bracketvar}.

Specializing to the area operator, for which  $f_1(v_1)=0=f_1(v_0)$,  this expression simplifies to
\beqn\label{lom}
\fL'_{\hat f_2} m_{v_0}^{v_1}[f_1]
&=&8\pi G m_{v_0}^{v_1}[[f_1,f_2]]-
[\Omega\pa_v[f_1,f_2]]^{v_1}_{v_0}
\\
&=&
\int_{v_0}^{v_1}\rd v
[f_1,f_2] C
-[[f_1,f_2]\pa_v\Omega]^{v_1}_{v_0}.
\eeqn
To compute the area-area bracket,  we now need to choose $f_2$ judiciously. We select vector fields satisfying
\beq\label{fcond}
f_1(v_0)=0\quad f_1(v_1)=0\quad \pa_v f_1(v_1)=1\qquad f_2(v_0)=0\quad f_2(v_2)=0\quad \pa_v f_2(v_2)=1.
\eeq
This implies
\beq
[f_1,f_2](v_0)=0,\qquad [f_1,f_2](v_1)=-f_2(v_1).
\eeq
We plan to explore further elsewhere the space of such test functions and the comparison between it and the Schwartz space of rapidly decreasing functions, \cite{treves2006topological}. 

Using this and the result in \eqref{lom}, we arrive at
\beq
\left\{ m_{v_0}^{v_1}[f_1],m_{v_0}^{v_2}[f_2]\right\}\hat{=}-8\pi G (\pa_v\Omega[f_1,f_2])(v_1)=8\pi G (\pa_v\Omega f_2)(v_1),
\eeq
which, restoring the cut dependency, and using $v_2 \geq v_1$,  reads
\beq
\left\{ m_{v_0}^{v_1}[f_1](z_1,\bz_1),m_{v_0}^{v_2}[f_2](z_2,\bz_2)\right\}\hat{=}-8\pi G (\pa_v\Omega[f_1,f_2])(x_1)\delta^{(2)}(z_{12}),
\eeq
and, given the conditions spelled out in \eqref{fcond},
\beq
m_{v_0}^{v_1}[f_1](z_1,\bz_1)\;\hat{=}\;\frac{1}{8\pi G}\Omega(x_1),\qquad m_{v_0}^{v_2}[f_2](z_2,\bz_2)\;\hat{=}\;\frac{1}{8\pi G}\Omega(x_2).
\eeq
Recalling that the expansion $\pa_v \Omega=\theta\Omega$ is the corner charge aspect for a null translation, we see that the Poisson bracket of two area operators is a translation, on shell. 

While we performed this computation at the classical level, using methods that we will develop in this paper, we expect its quantum version to instruct us about the area fluctuation on finite-distance hypersurfaces. In a later section, we will consider similar physics in the celestial context.

\subsection{Primed Diffeomorphisms}\label{prim}

Let us review the derivation of the primed phase space performed in \cite{Ciambelli:2023mir}. The symmetries of a null hypersurface are the diffeomorphisms of $\cN$ and the rescaling of $\ell$, which we refer to as a boost. The former act on the phase space via the ordinary Lie derivative. We focus on time reparametrizations, which is the symmetry generated by the Raychaudhuri constraint. For a vector field $f\ell$, one has
\beq
\fL_{\hat{f}}q_{ab}=\cL_f q_{ab},\qquad
\fL_{\hat{f}} \sigma_a{}^b =   \pa_v( f\sigma_a{}^b),\qquad \fL_{\hat{f}}\ell=-\ell(f)\ell\\
\fL_{\hat{f}}\mu=\pa_v(f\mu),\qquad
\fL_{\hat{f}}\Omega=f\pa_v\Omega,\qquad
\fL_{\hat{f}}\theta=\pa_v(f\theta),
\eeq
where $\fL_{\hat{f}}$ and $\cL_f$ are the phase space and spacetime ($\cN$) Lie derivatives, respectively.
The Carrollian structure (and in particular the condition $\ell^aq_{ab}=0$) is also preserved under boosts, which are rescalings of the Carrollian vector field $\ell$ generated by the infinitesimal parameter $\lambda$,
\beq
\fL_{\hat{\lambda}}q_{ab}=0,\qquad
\fL_{\hat{\lambda}} \sigma_a{}^b =\lambda\sigma_a{}^b,\qquad \fL_{\hat{\lambda}}\ell=\lambda\ell\\
\fL_{\hat{\lambda}}\mu=\lambda\mu+\ell(\lambda),\qquad
\fL_{\hat{\lambda}}\Omega=0,\qquad
\fL_{\hat{\lambda}}\theta=\lambda\theta\label{ltr}
.
\eeq
Combining these two phase space actions, we can consistently preserve the condition $\delta\ell=0$. This is done choosing $\ell=\pa_v$ and $\lambda_f=\pa_v f$, and introducing the primed phase space Lie derivative
\beq
\fL'_{\hat f}=\fL_{\hat f}+\fL_{\hat\lambda_f},\qquad  \fL'_{\hat f}\ell=0.
\eeq
Therefore, on the primed phase space $\ell$ is a background structure, and this will be important throughout this manuscript. The action of the symmetries on the primed phase space is collected in (\ref{trs})--(\ref{mutr}).

We remarked above that the constraint $C_f$ transforms $\mu$ as a connection rather than as just a 1-form, due to the shift in the boost action \eqref{ltr},
\beq
\{C_f,\mu\}=8\pi G\,\Big( \pa_v(f\mu)+\pa_v^2f\Big).
\eeq
There is a function on phase space which does treat $\mu$ as a 1-form. Introducing the current
\beq\label{cur}
\dC_f=\int_{\cN}\ve^{(0)}_{\cN}f \dC,
\eeq
we have
\beq\label{Tmuf}
\{\dC_f,\mu\}=8\pi G\,   \pa_v(f\mu).
\eeq
One finds that $C$ and $\dC$ are closely related, and in fact\footnote{These tensors are actually densities, ready to be integrated with the coordinate measure. This makes all the field dependencies appearing in the associated currents explicit.}
\beq\label{Tfirst}
C = \pa_v^2\Omega+\dC,\qquad  \dC:=-\mu\pa_v\Omega+\Omega(\sigma_{a}{}^{b}\sigma_{b}{}^a+8\pi G T^{\mathrm{mat}}_{vv}).
\eeq
Whereas $C$ vanishes on-shell in the classical theory, $\dC$ does not, at least in general. Nevertheless, using \eqref{Calg}, one can prove (see Appendix \ref{class}) that $\dC$ satisfies the same classical algebra,
\beq\label{Talg}
\{ \dC_f,\dC_g\}=- 8\pi G\,\dC_{[f,g]},
\eeq
and we note  that
\beq
8\pi G\,M_f=\dC_f+\int_{\cN}\ve^{(0)}_{\cN}\Omega\, \pa_v^2f.
\eeq
To summarize, $\dC_f$ acts differently on phase space than does $C_f$. In fact, $\dC$ acts by the ordinary `un-primed' Lie derivative $\fL_{\hat f}$, while $C$ acts with $\fL'_{\hat f}$. We recall that $\fL'_{\hat f}$ consists of a diffeomorphism combined with a boost chosen to preserve the null vector $\ell$. This can be understood as an `improvement' or `covariantization' of the diffeomorphism. In covariant phase space formalism \cite{Freidel:2016bxd}, this can also be thought of as an anomaly, whereby the symmetry (here the primed diffeomorphism) acts in phase space differently than one would expect from the spacetime perspective (where one might naively have taken $\mu$ to be a 1-form). Thus, the covariant phase space anomaly is nothing but the fact that $\mu$ is a boost connection rather than an ordinary 1-form. We stress again that $\dC$ does not generate a local symmetry of our system in general; rather, it is $C$ that does so.

There is also a close analogy here to familiar 2d conformal field theories that will play a central role in our reasoning. The primed diffeomorphism is the analogue of conformal transformations. The latter are a combination of a diffeomorphism and a certain Weyl transformation. It is a familiar property of 2d CFTs that conformal symmetry is anomalous in the sense of there being a central charge. This anomaly arises because it is impossible to introduce a regulator without violating Weyl invariance.\footnote{In (critical) string theory, one insists on gauging Weyl invariance. So, one requires the central charge to vanish. In our context, we do not have the 2d metric of string theory, but one can think of the  Carroll structure as playing an analogous role. Our construction of the quantum theory can be thought of as more closely related to non-critical strings.} We will claim that in the quantum algebra of the Raychaudhuri constraint, there is a (field-independent) central charge. It is crucial to realize that, just as in 2d CFTs in general, this arises not because diffeomorphisms are anomalous, but because an anomalous `global' symmetry is present. In our context, we will interpret the theory in terms of a {\it chiral} CFT (per point on a cut), and the central charge will arise through an anomaly in the rescaling symmetry. As far as this chiral CFT is concerned, this rescaling looks just like a Weyl symmetry. The presence of the central charge instructs us to think that only diffeomorphisms should be gauged in the quantum theory, not the rescaling.

\subsection{Conformal Time and Caustics}\label{conftime}

The symplectic structure on the primed phase space is given by \eqref{ocan2}. In \cite{Ciambelli:2023mir}, we showed that this can be rewritten in terms of fields that are dressed to diffeomorphisms and the constraint emerges as a variable conjugate to a dressing field. 
The dressing field is a scalar field  $V(v)$ which acts as a clock and is defined through the differential equation
\be \label{Vdef}
\pa_v^2 V =\mu \pa_vV.
\ee 
In the realm where $v\to V(v, z,\bar{z})$ is invertible it can be used to define dressed fields, which are gauge invariant. For instance
\be 
\tilde{\phi} = \phi \circ V^{-1} , 
\qquad 
\tilde{\sigma}_{ab} = (\pa_v V)^{-1} (\sigma \circ V^{-1}),
\ee 
where $\phi$ denotes scalar matter fields and tilde refers to the dressing. By construction, the dressed fields are invariant under time reparameterization, infinitesimally
\beq\label{vestito}
\fL'_{\hat f}\tilde\sigma_{ab}=0.
\eeq

The symplectic potential of the gravitational theory can be conveniently written in terms of the dressing variables as the sum of a bulk and a corner term
$\Theta^{\mathsf{can}}=\Theta^{\mathsf{can}}_{\cN}+ \Theta^{\mathsf{can}}_{\pa\cN}$, \cite{Klinger:2023auu},
\bee
\Theta^{\mathsf{can}}_{\cN} &=\frac{1}{8\pi G}
\int_{\cal{N}}\tilde\ve_{\cal{N}}^{(0)}
\left(\tfrac12 \tilde\Omega \tilde \sigma^{ab} \delta \tilde{q}_{ab}
+ \tilde{C}   \varpi )\right),\qquad
\Theta^{\mathsf{can}}_{\pa\cN}=\frac{1}{8\pi G}\int_{\pa\cN} \tilde\ve_{\cal{\cC}}^{(0)}\left( \varpi \partial_V \tilde \Omega - \partial_V  \varpi \tilde\Omega \right).
\eee 
In this expression $\varpi  $ denotes the Maurer-Cartan form
\be 
\varpi:= \delta V \circ V^{-1}.
\ee 
These formulas tell us that $\varpi$ represents the variable that is conjugate to the constraint from the hypersurface perspective. Furthermore, once the constraint is imposed, there is a residual corner symplectic potential in which  $\varpi$, representing the translation time, 
is conjugate to the expansion on the corner, while   $\pa_V\varpi$, representing the boost time, is conjugate to the area. These are the physical gauge charge terms, and they are given by translation and boost because these represent the generators of the residual corner symmetries (represented in local coordinates by  $\pa_v$ and $v\pa_v$, respectively) supporting non-trivial gauge charges. We recall that in this manuscript we use the terms corner and cut interchangeably. 

Note that the values of $V$ and $\pa_v V$ at a corner represent initial values, necessary for the integration of \eqref{Vdef}. Including them in the phase space is necessary to perform the dressing. These corner modes are Goldstone modes or edge mode degrees of freedom, necessary to construct gauge invariant observables that possess well-defined commutation relations.

While this discussion is relevant for a single isolated corner, we are interested in corners along the null hypersurface. In fact, one can view \eqref{Vdef} as the key equation that relates corners to each other. Indeed, integrating it once, we find (we suppress the $z,\bz$ dependence in what follows)
\beq
\pa_v V(v)=\pa_v V(v_0)\, \exp\Big(\int_{v_0}^v \rd v'\mu(v')\Big).
\eeq
So the value of $\mu$ determines the dressing time on any cut at constant $v$.

As seen in \cite{Ciambelli:2023mir}, the dressing time plays a special role because it ensures that the energy flux is positive. It also allows for a definition of entropy in a dynamical regime. It is, however, not the only time one can extract from the spin-$0$ sector with special features. Furthermore, as we will see in Sections \ref{sec:CFT} and \ref{qutime}, quantization depends on the choice of time, and therefore, it is helpful to understand the different options at our disposal.

The basic mechanism used to construct the dressing time is that $\mu$ transforms under primed diffeomorphisms as a connection. This key property is true for any combination 
\be 
\mu_{(a)} := \mu + a \theta,
\ee 
and we can define a physical time $V_{a}$ by demanding that $\pa_v^2 V_{a} = \mu_{(a)} \pa_vV_{a}$. From the connection $\mu_{(a)}$ we can construct the analog of a stress tensor. Defining
\be \label{defnu}
\nu_{(a)}:= \pa_v\mu_{(a)} -\tfrac12 \mu_{(a)}^2,
\ee 
we have that under the primed action
\be 
\fL'_{\hat{f}}\mu_{(a)} = \pa_v(f\mu_{(a)}) + \pa_v^2f, 
\qquad
\fL'_{\hat{f}}\nu_{(a)} = f\pa_v\nu_{(a)} + 2\pa_v f \nu_{(a)} 
+ \pa_v^3 f.
\ee 
The finite transformation, calling $v'(v)$ the finite diffeomorphism generated by $f$, is indeed given by
\beq
\nu'_{(a)}(v')=\left(\pa_v v'\right)^{-2}\Big(\nu_{(a)}(v)-\{v';v\}\Big),
\eeq
where we introduced the Schwarzian derivative $\{v';v \} = \pa_v\left(\frac{\pa_v^2v'}{\pa_vv'}\right)- \frac12\left(\frac{\pa_v^2v'}{\pa_vv'}\right)^2 $.

From these transformations, we see that the algebra of transformations that preserves the condition $\mu_{(a)}=0$ is the affine algebra generated by $(\pa_{V_{a}}, V_{a}\pa_{V_{a}})$  \cite{Chandrasekaran:2018aop}. Moreover, the algebra of transformations that preserves the condition $\nu_{(a)}=0$ is the conformal algebra $\mathfrak{sl}(2,\RR)$ generated by  $(\pa_{V_{a}}, V_{a}\pa_{V_{a}}, V_{a}^2\pa_{V_{a}})$, which indeed has vanishing Schwarzian derivative.
For each choice of $a$, we thus have a choice of time $V_{a}$, which is well defined up to conformal transformation. Later we will see that the appearance of an $\mathfrak{sl}(2,\RR)$ is at the core of the quantization of the phase space. 

Recall that $\mu_{(a)}=\mu+a\theta=\kappa+(a+\frac12)\theta$. Setting it to zero, three choices of time are particularly relevant: $\mu_{(a)}=0$ for $a=-1/2$ corresponds to the affine time, $\mu_{(a)}=0$ for $a=0$ is the dressing time, while $\mu_{(a)}=0$ for $a=-3/2$ is the conformal time $V_c$ that will play a key role at the quantum level.
What is unique about the conformal time is that  $\mu_{(c)}:= \kappa -\theta$ is conformally invariant under bulk conformal transformations. 
Indeed, under a conformal transformation $g_{ab}\to e^{2\varphi} g_{ab}$, which  implies $\Omega\mapsto e^{2\varphi}\Omega$, we have that  $\kappa\mapsto\kappa+2\ell(\varphi)$ and $\theta\mapsto\theta+2\ell(\varphi)$. Notice that this conformal transformation does not act on $\ell$ \cite{Hopfmuller:2018fni}, which remains a background structure. This is important for the quantization proposed in this manuscript.
Therefore, only for the conformal time $V_c= V_{-3/2}$ the connection $\mu_{(c)}$ is invariant under this symmetry. Given that we will find a chiral CFT on the null time, this bulk conformal transformation is expected to play a prominent role, as it instructs us about the conformal rescalings of the $2$-planes normal to corners in the bulk. 

A general null hypersurface develops caustics, where $\Omega\to 0$. Near a caustic, we can use the area element as time. This is the areal time that we studied in \cite{Ciambelli:2023mir}. The Raychaudhuri equation implies that the relationship between dressing time and area is 
\beq\label{solRaych}
\pa_\Omega \bar{V}=
\exp\int^\Omega_0 \D\Omega'\,\Omega'(\bar\sigma^2+8\pi GT^{\mathsf{mat}}_{\Omega\Omega})(\Omega').
\eeq
where $\bar{\phi}$ refers to the field $\phi$ expressed in areal time.
From this, we see that near a caustic, where $\Omega'\to 0$, the dressing time behaves as in flat space since we have $\bar{V}=\Omega$. 
This also means that near a caustic the affine time is $\sqrt{\Omega}$ and the conformal time is its inverse $-1/\sqrt{\Omega}$. 

To gather intuition, we can explicitly construct these different notions of time for a null cone $t-r=0$ in flat space. The induced null metric is $ \rd s^2 \stackrel{\cN}{=}\tfrac14 v^2(\rd \theta^2+\sin^2\theta\rd \phi^2)$.
The affine time is the advanced time $v=t+r$. We can evaluate the surface tension associated with a generic vector field $\ell=e^{\alpha(v,\theta,\phi)} \pa_v $ to be 
\be 
\mu_{(a)} = e^{\alpha} \left( \pa_v \alpha + \frac{(2a+1)}{v}\right). 
\ee
From this expression, we readily see that requiring to be in affine time, $\mu_{(a)}=0$ with $a=-1/2$, we have $\alpha=0$ and thus $V_{-1/2}=v$. Similarly, the dressing time condition $\mu_{(a)}=0$ for $a=0$, gives $\alpha=-\log{v}$, which implies $\ell=v^{-1}\pa_v$ and thus $V_0=v^2/2$. The dressing time is exactly the area. Finally, the conformal time is achieved by setting $\mu_{(a)}=0$ with $a=-3/2$, which yields $\alpha=2\log{v}$. This implies $\ell=v^2\pa_v$ and thus $V_c = -1/v$. In general, we have, up to an overall constant, that 
\beq
V_a=\frac{v^{2(a+1)}}{2(a+1)}, \qquad \ell=  
\pa_{V_{a}}=v^{-(2a+1)}\pa_v.
\eeq
We can distinguish these times in two classes depending on how they behave at the caustic $v=0$. 

A caustic avoiding time (CAT) is defined to be a time for which $\Omega\to 0$ always implies $\theta\to 0$. While the analysis of caustics in full generality is a complicated task \cite{arnold1985singularities, arnold1990singularities, Gadioux:2023pmw}, in this paper we restrict our analysis to null vectors parametrized as $\ell=\pa_V$, and to caustics appearing at cuts $V=$cst.
For such caustics, CATs necessarily have that $\theta \propto \Omega^\gamma$ with $\gamma>0$ near the caustic. Using that $\theta=\pa_V \Omega /\Omega \sim \Omega^\gamma$, we can integrate this equation to
$V \propto \Omega^{-\gamma}$. Therefore we see that for this specific family of CATs the time range goes to infinity at the caustic.\footnote{A more general class of caustic is accessible if we use a general parametrization $\ell=e^\alpha(\pa_V + U^A \pa_A)$ and choose the caustic to be at $V=$cst. In this case, the time $V$ does not have to be infinite at the caustic as long as $e^\alpha\to 0$. 
It is only when we reabsorb this prefactor and the twist vector $U^A$ into the definition of time that caustic avoiding means infinite time. In general, the $V=$cst caustic is then mapped into  $V= T(z,\bz)$ and a new analysis is needed.} It is inextensible there, and this will be crucial for the quantization. To summarize, within our restrictive assumptions, caustic avoiding times are such that $\theta\to 0$ when $\Omega\to 0$, and $V\to \infty$ at the caustic. This happens in flat space when $a < -1 $. In this case we see that the infinitesimal interval $v=(0,\epsilon]$ near the caustic is mapped to $V_a=\left(-\infty, \frac{\epsilon^{2(a+1)}}{2(a+1)}\right] $.\footnote{The limiting case $a=-1$ is when $V_{-1}= \beta^{-1} \ln v$ is the thermal, boost or Rindler time with vector field $\ell= \beta v\pa_v$.} This discussion is valid only for null cones in flat space so far. However, due to the argument presented around \eqref{solRaych}, we know that near a caustic the flat space description is a good approximation of the general geometry. This seems to indicate a universal behaviour of geometry near caustics, that deserves further exploration. Here, we conclude that times $V_a$ for $a<-1$ are inextensible CATs.

Among these values we have $a=-3/2$, which corresponds to the choice of conformal time. We will argue next that this value of $a$ is special. It corresponds to a vector field that behaves near the caustic as 
\be 
\pa_{V_c}= v^2 \pa_v,
\ee 
where $v$ denotes the affine time.
In flat space, the choice of conformal time can be shown to arise naturally for the description of CFTs on null cones because of the following facts. First, a Rindler plane is mapped onto a null cone by the conformal inversion map and second, the Rindler time vector field $\pa_v$ which describes the flat space vacuum on the Rindler plane is mapped to the conformal time vector field $v^2\pa_v$ by the inversion 
(see \cite{Bousso:2014uxa, Bousso:2014sda}).
These essential properties firmly suggest that the conformal time is the right time to ground our quantization scheme. The time $V_c$ defined earlier is a natural \emph{background independent} candidate for a choice of time that possesses $\mathfrak{sl}(2,\mathbb{R})$ as a symmetry. The fact that such a background-independent definition exists is quite remarkable.

To conclude this section, we observe that it is crucial to perform quantization in an inextensible caustic avoiding time, for two reasons. The first is that for such times the value of $v$ on the portion of null hypersurface between caustics $\cN$ runs from $-\infty$ to $\infty$, which is important when using tempered distributions. The second reason is that CATs remove the symplectic flux at the boundary: both $\theta$ and $\Omega$ vanish at the caustic in these times. This makes the flux \eqref{sympf} vanish there, and so we have a well-defined classical phase space, which is why we can propose a consistent quantization.

\section{Quantum Gravity CFT}\label{sec:CFT}

Classically, the vanishing of the Raychaudhuri constraint  should be understood as expressing diffeomorphism invariance on the null hypersurface. As such, together with ultralocality, it is natural to interpret it as a chiral stress tensor in a 2d CFT. Furthermore, we suppose that it is promoted to an operator in the quantum theory,
\beq\label{qC}
\hat C(x)=\frac{C(x)}{4G\hbar},
\eeq
where the normalization will ensure that the central charge is dimensionless. It is important to emphasize that what we are describing here is not a celestial CFT on a cut (which would give rise to a 2d Euclidean CFT only in 4d gravity), but a CFT per point on a cut. By focusing our attention on a null hypersurface, we are seeing the chiral (light-cone) half of what presumably is a full 2d CFT on the Lorentzian plane normal to a cut.\footnote{In 2d the light cones are simply a pair of lines and the CFT described here is similar to that discussed by Solodukhin \cite{Solodukhin:1998tc}. } From this perspective, there is a 2d CFT lurking in quantum gravity everywhere, in {\it any} spacetime dimension. 

Let us take this proposal seriously. A robust prediction is that there is an anomaly in the quantum theory that shows up, for example, in the operator product expansion (OPE)\footnote{While formally $u,v$ are analogous to $w,\bar w$ in Euclidean signature, they are real and not complex variables. Moreover we here have Diff$(\RR)$ symmetries and not Diff$(S^1)$. We can take this into account by keeping track of the contour via the $i\epsilon$ prescription and taking $\frac{1}{v_{12}-i\varepsilon}$ in place of $\frac{1}{z_{12}}$. In the complex case the $i\epsilon$ prescription is implicit in the fact that one assumes the radial quantization $Im(z_{12})<0$ in the product of fields. Further discussion of related points will follow. }
\beq\label{TTope}
\hat C(x_1) \hat C(x_2)\sim \left(\frac{c N}{2(v_{12}-i\epsilon)^4}\pmb 1+\frac{2 \hat C(x_2)}{(v_{12}-i\epsilon)^2}
+\frac{\pa_{v_2}\hat C(x_2)}{v_{12}-i\epsilon}\right) \delta^{(2)}(z_{12}).
\eeq
The total central charge is $c N$, where $c$ is the central charge on each null generator, and $N$ counts the number of null rays. The number of null rays is in one-to-one correspondence with the number of points on the chosen cut $\cC$. The  central charge $c$ depends on the field content of the theory. In the following sections, we evaluate it perturbatively in $G$ and show in that regime that
\be \label{cc}
c=4 + c^{\mathrm{mat}},
\ee
where $c^{\mathrm{mat}}$ is the matter central charge.
This shows that the central charge is {\it non-zero}.
Determining the non-perturbative value of $c$ is a task that we leave for future works --- it is not straightforward, since we have an interacting theory. Indeed, recall that the Raychaudhuri constraint has the form
\beq
C=\pa_v^2\Omega-\mu\pa_v\Omega+\Omega (\sigma_{a}{}^{b}\sigma_{b}{}^a+8\pi G T^{\mathrm{mat}}_{vv}).
\eeq
Splitting the constraint into its spin-0, -2, and matter parts, as done classically in \cite{Ciambelli:2023mir}, is a challenging task, because the different sectors do not satisfy a closed algebra separately. Nonetheless, an important repercussion of promoting the entire Raychaudhuri constraint to a quantum operator is that the spin-0 sector is also quantized, and thus we have a quantum geometric interpretation of the area operator. 
Note that the stress tensor is at most cubic in the elementary fields. While the spin-$0$ piece is  quadratic in the field, the cubic coupling of the spin $0$ to the spin-$2$ and matter fields is universal, simply involving the area form $\Omega$ as an overall coupling. This leads to a physical picture where the spin-$0$ fields act as universal fluctuating elements creating quantum noise for the matter fields. This is  in agreement with the treatment of Parikh et al. \cite{Parikh:2020fhy,Bak:2023wwo} on the effects of  gravitational perturbations as a source of stochastic noise that can be modeled through a Langevin equation and the Feynman-Vernon influence functional. In our case, the source of noise is due to quantum fluctuations of the spin-$0$ field.

In (non-critical) string theory, it is well known that the anomaly is in the worldsheet Weyl symmetry, with regulators preserving the worldsheet diffeomorphism invariance readily available. In the present case, we should expect hypersurface diffeomorphism invariance to be non-anomalous; it is the internal rescaling symmetry that is anomalous, and the anomaly shows up in the physics of the quantum Raychaudhuri constraint because it is the Noether constraint of the primed diffeomorphisms, which are a combination of diffeomorphisms and a particular internal rescaling. This is directly analogous to string theory, in which conformal transformations are a combination of particular diffeomorphisms and Weyl transformations, the conformal anomaly then arising because of the Weyl anomaly. Note that here, as already stressed, while we do not have a 2d metric theory, the internal rescaling is indistinguishable from a Weyl transformation because we are seeing only a chiral half of a 2d theory, the role of the worldsheet metric replaced by the Carroll structure. 

\subsection{Spin-0 \texorpdfstring{$\beta\gamma$}{bg} CFT}\label{bg}

A simplified setup that one can consider is the absence of matter and radiation. In this case, the Raychaudhuri constraint and $\dC$ as defined in \eqref{Tfirst} read
\beq\label{Ray0}
C=\pa_v^2\Omega-\mu\pa_v\Omega, \qquad \dC=-\mu\pa_v \Omega.
\eeq
The only fields are $\Omega$ and $\mu$, satisfying the Heisenberg algebra ($x_{12}=x_1-x_2$)
\beq\label{PBs}
\{\Omega_1,\mu_2\}=8\pi G\,\delta^{(3)}(x_{12}),\qquad \{\Omega_1,\Omega_2\}=0\qquad \{\mu_1,\mu_2\}=0.
\eeq

This system can be canonically quantized: $\{\cdot,\cdot\}\to \frac{1}{i\hbar}[\hat\cdot,\hat\cdot]$. The Poisson bracket then becomes the commutator
\beq
[\hat\mu_1,\hat\Omega_2]=-i8\pi G\hbar\; \delta^{(3)}(x_{12}),
\eeq
where we have promoted the classical  fields to quantum operators. This holds if we choose a caustic avoiding time $v$ extending from $-\infty$ to $+\infty$ on the null hypersurface, such as those discussed in Section \ref{conftime}.
The importance of choosing a time for quantization will be thoroughly discussed in the next section.
Indeed this analysis can only be performed with specific choices of time (which corresponds in constraint quantization to particular choices of gauge).  The central charge will play an important role in such discussions, through the Schwarzian derivative appearing in the transformation law of the stress tensor.  

The fields $\hat\mu$ and $\hat\Omega$  depend on the coordinates $(z,\bz)$ on the cut. However, thanks to the ultralocal nature of null physics, the algebra of operators tensor-factorizes into the product of algebras for each null generator. In other words, the Poisson bracket in the classical case or the commutator in the quantum case between fields at different points on the cut vanishes. This implies that we can focus only on the $v$-dependency, that is, we can work per point on $\cC$. 

To evaluate the OPE of $\hat\Omega$ and $\hat\mu$, we introduce the normal direction $\pa_u$ to the null hypersurface $\cN$, in the bulk, observe that the constraint satisfies $\pa_u C=0$, and furthermore that the fields are chiral. We can then derive the OPE on each null generator using standard techniques, see e.g. \cite{Green:1987sp},
\beq\label{mphi}
\hat\mu_1\hat\Omega_2\sim -\frac{4G\hbar}{v_{12}-i\epsilon}.
\eeq
Here and in the following we use the shortcut notation $\Omega_1=\Omega(x_1)=\Omega(v_1,z_1,\bz_1)$.
We stress again that this analysis is performed in a time $v\in \mathbb{R}$ on $\cN$. Furthermore, from \eqref{PBs}, we also find that $\hat\Omega$ and $\hat\mu$ have only analytic correlations, 
\beq
\hat\Omega_1\hat\Omega_2\sim 0, \qquad \hat\mu_1\hat\mu_2\sim 0.
\eeq
This is due to the absence of spin-$2$ and matter interactions, in the context of this section. We will perturbatively reinstate those in the next section, and indeed the area perturbation will acquire non-analytic correlations.

Since we have a Heisenberg algebra, it is straightforward to introduce a notion of normal ordering, and we write the quantum Raychaudhuri constraint \eqref{qC} as 
\beq
\hat C=\frac1{4G\hbar} (\pa_v^2\hat \Omega-
\colon\!\hat\mu\pa_v\hat\Omega\!: ).
\eeq

We then remark that this Raychaudhuri constraint and the OPE \eqref{mphi} are exactly the stress tensor and OPE of a (Lorentzian) curved bosonic $\beta\gamma$ CFT, see \cite{Losev:2005pu, Nekrasov:2005wg} as well as \cite{Polchinski:1998rq, Kapustin:2006pk}, with conformal weights $h_\beta=1$ and $h_{\gamma}=0$, which corresponds to $\lambda=1$ in the language of \cite{Polchinski:1998rq}. Indeed, the $\beta\gamma$ OPE is 
\beq
\hat\beta_1\hat\gamma_2\sim-\frac{1}{v_{12}-i\epsilon},
\eeq
and the curved (aka twist) stress tensor for $\lambda=1$ is
\beq\label{twbg}
\tilde T^{\beta\gamma}=\frac12 \pa_v^2\log \omega(\hat\gamma)-:\!\hat\beta\pa_v\hat\gamma\!:,
\eeq
where $\omega(\hat\gamma)$ is the volume element of the holomorphic top degree form in target space.
This leads us to the identification\footnote{This identification is further suggested by the fact that $\hat\beta$ is the component of a 1-form \cite{Nekrasov:2005wg}, and so is $\hat\mu$.}
\beq
\hat\beta=\hat\mu,\qquad \hat\gamma=\frac{\hat\Omega}{4G\hbar},
\eeq
together with $\omega(\hat\gamma)=e^{2\hat\gamma}$ and $\tilde T^{\beta\gamma}=\hat C$.

Therefore, the surface tension operator $\hat\mu$ and the area element in Planck units play, respectively, the role of  $\hat\beta$ and $\hat\gamma$ in this chiral CFT. It is interesting to remark that the correct proportionality factor for the relationship between $\hat\gamma$ and $\hat\Omega$ is exactly the Bekenstein-Hawking numerical factor, suggesting that $\hat\gamma$ can be interpreted as an infinitesimal entropy operator.

Using the OPE \eqref{mphi} and the stress tensor, we can easily compute the operator product expansion 
\beq\label{quan}
\tilde T^{\beta\gamma}(v_1) \tilde T^{\beta\gamma}(v_2) \sim \frac{1}{(v_{12}-i\epsilon)^4}+\frac{2 \tilde T^{\beta\gamma}(v_2)}{(v_{12}-i\epsilon)^2}
+\frac{\pa_{v_2}\tilde T^{\beta\gamma}(v_2)}{v_{12}-i\epsilon}.
\eeq
The central term comes from the double contraction of $\hat\beta_1, \hat\beta_2$ with $\pa_v\hat\gamma_2$ and $\pa_v\hat\gamma_1$. 

Comparing with \eqref{TTope}, and using $N=1$ since we are focusing on a single null ray, this shows that the  spin-0 central charge on each null generator is $c=2$, which is indeed the expected result given that $c=3(2\lambda-1)^2-1$ and here we have $\lambda=1$. 

We remark that $\dC$ in \eqref{Ray0} is the untwisted version of $\tilde T^{\beta\gamma}$. Indeed, if one defines
\beq\label{untwbg}
T^{\beta\gamma}=-\frac1{4G\hbar}  \colon \! \hat\mu \pa_v \hat\Omega \colon,
\eeq
one also gets
\beq\label{quanT}
T^{\beta\gamma}(v_1) T^{\beta\gamma}(v_2) \sim \frac{1}{(v_{12}-i\epsilon)^4}+\frac{2 T^{\beta\gamma}(v_2)}{(v_{12}-i\epsilon)^2}
+\frac{\pa_{v_2}T^{\beta\gamma}(v_2)}{v_{12}-i\epsilon}.
\eeq
Therefore, we have shown that $T^{\beta\gamma}$ satisfies the same OPE, and it is related to $\tilde T^{\beta\gamma}$ via 
\beq
\tilde T^{\beta\gamma}=T^{\beta\gamma}+\frac{\pa_v^2 \hat\Omega}{4G\hbar}.
\eeq

If one reintroduces the spatial dependence, thanks again to ultralocality, this simply amounts to adding  a $\delta^{(2)}(z_{12})$ in the $\beta\gamma$ OPE. The consequences are however dramatic, as one readily obtains
\beq
\tilde T^{\beta\gamma}(x_1)\tilde T^{\beta\gamma}(x_2) \sim \left(\frac{\delta^{(2)}(0)}{(v_{12}-i\epsilon)^4}+\frac{2 \tilde T^{\beta\gamma}(x_2)}{(v_{12}-i\epsilon)^2}
+\frac{\pa_{v_2}\tilde T^{\beta\gamma}(x_2)}{v_{12}-i\epsilon}\right)\delta^{(2)}(z_{12}),
\eeq
where we used $\delta^{(2)}(z_{12})\delta^{(2)}(z_{12})=\delta^{(2)}(0)\delta^{(2)}(z_{12})$. The $\delta^{(2)}(0)$ counts the `number' of points on the cut, it plays the role of $N$ and  indicates that  we have $N\to \infty$, leading to an infinite central charge. This simultaneously establishes two facts: first, it shows that this central charge indeed counts the fields $\hat\Omega$ and $\hat \mu$ ($c=2$) times the number of null generators. As such, it truly counts the geometric degrees of freedom. Secondly, it shows the universal nature of this divergence, simply given by the infinitely many null generators. This was already noticed in \cite{Wall:2011hj} for the matter sector. We have given here a quantum derivation for the spin-$0$ sector. One could have anticipated this from the ultralocal nature of the commutators (brackets). In Section \ref{mol}, we will propose a different representation for the area operator $\hat\Omega$, which entails a finite central charge. 

\subsection{Perturbative Gravity CFT}

We wish now to reintroduce the radiative and matter degrees of freedom. We do so perturbatively, in the weak gravity regime. In this framework, we derive the perturbative Poisson brackets and canonically quantize the system. While for non-expanding backgrounds we can confirm and generalize the results of the previous section, for generic expanding backgrounds we offer some preliminary explorations of the quantization procedure.

We expand fields in the limit $G\to 0$, using the expansion parameter $\epsilon:= \sqrt{8\pi G}$.
Given \eqref{bp}, we encode the perturbative spin-$2$ gravitons in the expansion of the Beltrami differentials
\beq
\zeta=\epsilon X + \cO(\epsilon^2).
\eeq
The spin-$0$ data define the background, and the matter and graviton dynamics only backreact on the background at order $\epsilon^2$. Thus we have the  expansion
\beq\label{bck}
\Omega=\Omega_{\B} +\epsilon^2 \omega+\cO(\epsilon^3), \qquad \mu=\mu_0 +\epsilon^2 \mu_{II}+\cO(\epsilon^3),
\eeq
where $\Omega_{\B}$ denotes the background area form and 
where we called the quadratic order $\omega$ to single it out for future purposes. We will see that it represents a null version of the Newtonian potential. Its fluctuations encode the backreaction of matter and gravitons onto geometry, and that is why it contributes at the second order (that is, at order $G$) in the perturbative expansion. 

With these expansions we can compute the leading order of the spin-$2$ part of the constraint
\beq\label{T2pert}
\Omega\sigma_{a}{}^{b}\sigma_{b}{}^a=2\epsilon^2 \Omega_{\B}\pa_v X\pa_v \bX+\cO(\epsilon^3).
\eeq
Therefore, since the matter stress tensor couples without extra factors of $G$, inserting these expansions into the Raychaudhuri constraint we find
\beq
C=C_B +\epsilon^2 \bC+\cO(\epsilon^3),
\eeq
with
\beq
C_B=\pa_v^2\Omega_{\B}-\mu_0\pa_v\Omega_{\B},\label{CB}\qquad
\bC =\pa_v^2\omega-\mu_0 \pa_v\omega-\mu_{II}\pa_v\Omega_{\B}+\Omega_{\B}\left(2\pa_v X\pa_v\bX+  T^{\mathrm{mat}}_{vv}\right).\label{C2}
\eeq

From \eqref{mutr}, we  have that $\mu$ transforms anomalously as a connection. An important conceptual step is that this can be used to reabsorb its perturbation $\mu_{II}$ as a time redefinition. This  is ultimately the rationale for having a well-posed perturbative setting around a time-dependent background.
Therefore, introducing a time perturbation $v_{II}(x)$ satisfying  
\be \label{pertDT}
\pa_v(\pa_v+\mu_0)v_{II} =\mu_{II},
\ee 
we can redefine time to be $v'=v +\epsilon^2 v_{II} + \cO(\epsilon^3)$, such that $\mu_{II}(x')=0$. In practice, one would have to impose this gauge  to all orders in order to have a well-defined perturbative expansion. Here, we will confine our attention to the expansion up to second order, for which the conditions spelled out above suffice. 
Given this gauge condition, we drop the subscript in $\mu_0 \rightarrow \mu$, and work in time $v'$ from now on. Given the discussion in \cite{Ciambelli:2023mir}, we call this time the {\it perturbative dressing time}.

The zeroth order Raychaudhuri constraint determines $\Omega_B$ to be a functional of $\mu$
\be\label{Omb}
\Omega^\mu_{\B}(x) = a(z,\zb) \left(\int_0^v \rd v' e^{\int_0^{v'}\rd v'' \mu(v'',z,\bz) }\right)  + b(z,\bz),
\ee
where $b(z,\zb)= \Omega^\mu_B(0,z,\bz)$ while $a(z,\zb)=\pa_v\Omega^\mu_B(0,z,\bz)$ are the initial data. Here we introduced the notation $\Omega^\mu_B$ to emphasize that $\Omega_B$ is a functional of $\mu$, on shell of $C_B=0$.
If we assume that $a=0$ then $\pa_v\Omega^\mu_B=0$:  in that case, the background structure is non-expanding, $\Omega^\mu_{B}$ is independent of $\mu$ and can be treated as a classical background structure. This is for instance the case for a perturbation around a black hole Killing horizon. This scenario has been studied in the recent literature starting with the seminal work of Wall \cite{Wall:2011hj}, and beyond \cite{Casini:2013rba, Faulkner:2024gst}.

If $a\neq 0$, there is now background expansion and $\Omega^\mu_B$ explicitly depends on $\mu$. This situation is suitable to describe e.g., a flat space null cone, in the time defined by $\mu$,\footnote{Recall that for a light cone we have that the affine time is simply $\sqrt{\Omega^\mu_B}$.}
and one can choose the reference point $v=0$ to be the tip of the cone. This second   case is very different since it means that the background is time-dependent and therefore the background area form is, in fact, an operator in the quantum theory. The fact that we can set up a perturbation scheme in this situation is new, and one of the main results of this manuscript. Notice that in such a regime $\Omega_{\B}^\mu$ plays the same role as the dressing time $V$ in the non-perturbative setting, as one can see comparing \eqref{Omb} with \eqref{Vdef}.

In perturbative dressing time, the second order
Raychaudhuri constraint \eqref{C2} becomes
\beq \label{CCFT}
\bC=\pa_v^2\omega-\mu \pa_v\omega+\Omega^\mu_{\B}\left(2\pa_v X\pa_v\bX+ T^{\mathrm{mat}}_{vv}\right)=\pa_v^2\omega+\bT,
\eeq
where we have introduced 
\be \label{TCFT}
\bT=-\mu \pa_v\omega+\Omega^\mu_{\B} \left(2\pa_v X\pa_v\bX+  T^{\mathrm{mat}}_{vv}\right).
\ee 

Our main goal is to promote the perturbative constraint \eqref{CCFT}
to a quantum operator and show that it satisfies \eqref{TTope}. We will see that  $\bT$, on the other hand, does not satisfy a closed algebra on expanding backgrounds. This is one of the main differences when treating expanding backgrounds: the spin-$0$ geometry is fully promoted to quantum operators, and it is not merely a spectator. Indeed, the area perturbation becomes a non-commutative field. Note furthermore that on shell of \eqref{CCFT}, one obtains
\be 
\pa_v^2\omega = -\bT.
\ee 
Quite remarkably, this identity looks like a saturation of the Quantum Null Energy Condition (QNEC) \cite{Bousso:2015mna}, provided we identify the area variation $\frac{\Delta A}{4 G} = 2\pi \int_{\cC}\ve_{\cC}^{(0)}\omega$ with the relative entanglement entropy.\footnote{The QNEC is $\langle \phi | T_{vv}(V,z,\bz)|\phi\rangle \geq \frac1{2\pi } \frac{\delta^2 S(\phi; V)}{\delta V(z,\bz)^2}$,  with $S(\phi; V)$ the  relative entanglement entropy   \cite{Bousso:2015mna,Faulkner:2016mzt,Kudler-Flam:2023hkl,Bousso:2015qqa}.}

Let us now give a proof that classically
\be \label{CCC}
\{\bC_f,\bC_g \}= -\bC_{[f,g]},
\ee 
where we used the same notation as in \eqref{Ccur} for the currents, $\bC_f = \int_{\cN}\ve_{\cN}^{(0)} f (\pa_v^2 \omega+ \bT)$. We will demonstrate this result in two complementary ways. In the first proof, we use that $\bC_f$ generates $\fL'_{\hat f}$ on the phase space. The second proof, in the next subsection, uses directly the perturbative Poisson bracket to evaluate the algebra.

Inserting the field perturbations in the symplectic two form \eqref{ocan2}, a long yet straightforward computation gives
\beqn
&\Omega^{\mathrm{can}}=\frac1{\epsilon^2}\int_{\cN}\ve_{\cN}^{(0)}\big(\delta\Omega_B\wedge \delta\mu+\epsilon^2(\delta\Omega_B\wedge (\pa_v \bX \delta X+\pa_v X\delta \bX)+\delta\omega\wedge \delta\mu&\nonumber\\
&+2\pa_v (\sqrt{\Omega_B}\delta X)\wedge \sqrt{\Omega_B}\delta \bX)+\cO(\epsilon^3)\big).&
\eeq
At this stage, a proliferation of phase space variables seems to occur, because $\Omega$ is expanded into two independent fields. However, imposing the background constraint $C_B=0$ results in \eqref{Omb}, which in turn implies $\delta\Omega_B^\mu\wedge \delta \mu=0$. This also renders the whole symplectic two-form finite in the $\epsilon\to 0$ limit, and yields
\beq\label{s2p}
\Omega^{\mathrm{can}}=\int_{\cN}\ve_{\cN}^{(0)}\big(\delta\Omega^\mu_B\wedge (\pa_v \bX \delta X+\pa_v X\delta \bX)+\delta\omega\wedge \delta\mu+2\pa_v (\sqrt{\Omega^\mu_B}\delta X)\wedge \sqrt{\Omega^\mu_B}\delta \bX\big)+\cO(\epsilon).
\eeq
This symplectic two-form stems from the canonical potential
\be 
\Omega^{\mathrm{can}} := \delta \Theta^{\mathrm{can}}, \qquad 
\Theta^{\mathrm{can}} = \int_{{\cN}}\ve_{\cN}^{(0)} \left(-
  \mu \delta \omega + \Omega_B^\mu (\pa_v X \delta \bX +\pa_v \bX \delta X)
  \right)  +\cO(\epsilon).
\ee 

Under the symmetry transformation 
\be 
\fL'_{\hat f} \mu = \pa_v(f\mu)+\pa_v^2f, \qquad \fL'_{\hat f} \omega =f\pa_v \omega, \qquad \fL'_{\hat f} X = f\pa_v X,
\ee 
we find that the background area form transforms as a primary field of weight $0$ if  $f$ and $\pa_vf$ vanish at the boundary of the null hypersurface. In other words
\be 
\fL'_{\hat f} \Omega_B^\mu= f \pa_v\Omega_B^\mu.
\ee 
Therefore, we can easily establish, using $\delta f =0$, that 
\bee 
\fL'_{\hat f} \Theta^{\mathrm{can}} &=- 
\delta \left(\int_{\cN}\ve_{\cN}^{(0)}  \pa_v^2 f \omega \right) 
+ \int_{\pa\cN}\ve_{\cC}^{(0)} f \left(-
  \mu \delta \omega + \Omega_B^\mu (\pa_v X \delta \bX +\pa_v \bX \delta X 
  \right), \cr
I'_{\hat f}  \Theta^{\mathrm{can}} 
&= \int_{\cN}\ve_{\cN}^{(0)}f \left(-\mu \pa_v \omega + 2\Omega_B^\mu \pa_v X\pa_v\bX \right).
\eee
Hence, using $f|_{\pa\cN}=0=\pa_v f|_{\pa\cN}$, we have 
\bee 
I'_{\hat f} \Omega^{\mathrm{can}}
= -\delta \int_{\cN} \ve_{\cN}^{(0)}
\left(
\pa_v^2 f \omega + f \bT
\right)= -
\delta \int_{\cN} \ve_{\cN}^{(0)}
f \left(
  \pa_v^2 \omega +  \bT
\right) = -\delta \, \bC_f.
\eee 
This shows that $\bC_f$ is the perturbative generator of gauge transformations  when $f$ and $\pa_vf$ vanish on the boundary of $\cN$. It thus  acts canonically on an arbitrary functional $O$ and satisfies the algebra
\be 
\{ \bC_f, O\} = \fL'_{\hat f} O, \qquad 
\{\bC_f,\bC_g \}= -\bC_{[f,g]},
\ee 
which is our desired result, eq. \eqref{CCC}.

Note that, generally, this would imply
\be 
\{\bT_f, \bT_g \}= -\bT_{[f,g]},
\ee 
if and only if the area perturbation is a commutative field, namely if it satisfies $\{\omega,\omega\}=0$. In general, one finds
\bee \label{TTal}
\{\bT_f, \bT_g \} =  -\bT_{[f,g]} + \int_{\cN}\ve_{\cN_1}^{(0)}\int_{\cN}\ve_{\cN_2}^{(0)}\pa_{v_1}^2 f_1 \pa_{v_2}^2 g_2
\{\omega_1,\omega_2\}.
\eee
In the following, we evaluate explicitly the perturbative Poisson bracket and show that $\{\omega,\omega\}$ vanishes only for non-expanding backgrounds, and therefore $\bT$ does not satisfy a closed algebra in general, only $\bC$ does. 

\subsubsection{Poisson Bracket}

We wish to extract the Poisson bracket in the perturbative regime. We relegate to Appendix \ref{A2} the inversion of the symplectic two form \eqref{s2p}, and report here the brackets among the elementary fields. First and foremost, introducing the notation
\beq\label{DO}
\Delta\Omega_{B12}^\mu=\Omega_{B1}^\mu-\Omega_{B2}^\mu,
\eeq
and the Heaviside function $\Theta(v_{21})$ as defined in \eqref{thhe},
one finds that the area perturbation $\omega$ satisfies
\beqn
\{\omega_1,\omega_2\}=-\int_{v_1}^{v_2}\rd v_3\int_{v_1}^{v_3}\rd v_4\int_{v_1}^{v_4}\rd v_5\frac{\pa_{v_3}\Omega^\mu_{B3}\Delta\Omega^\mu_{B51}}{2\sqrt{\Omega^\mu_{B4}\Omega^\mu_{B5}}}\left(\pa_{v_5}\bX_5\pa_{v_4}X_4+\pa_{v_5}X_5\pa_{v_4}\bX_4\right)\Theta(v_{21})\delta^{(2)}(z_{12}).\nonumber
\eeqn
Moreover, we find
\beqn
\{\omega_1,\Omega_{B2}^\mu\}&=&\Delta\Omega^\mu_{B 21}\Theta(v_{21})\delta^{(2)}(z_{12})\label{omOm}\\
\{\omega_1,X_2\}&=&-\int_{v_1}^{v_2}\frac{\pa_{v_3}X_3}{2\sqrt{\Omega^\mu_{B2}\Omega^\mu_{B3}}}\Delta\Omega^\mu_{B31}\rd v_3 \ \Theta(v_{21})\delta^{(2)}(z_{12})\\
\{\omega_1,\bX_2\}&=&-\int_{v_1}^{v_2}\frac{\pa_{v_3}\bX_3}{2\sqrt{\Omega^\mu_{B2}\Omega^\mu_{B3}}}\Delta\Omega^\mu_{B31}\rd v_3 \ \Theta(v_{21})\delta^{(2)}(z_{12})\\
\{X_1,\bX_2\}&=&-\frac{\Theta(v_{21}) \delta^{(2)}(z_{12})}{2\sqrt{\Omega^\mu_{B1}\Omega^\mu_{B2}}}.
\eeqn
A good consistency check is that we get
\beq
\{\omega_1,\mu_2\}=\pa_{v_2}\left(\frac{\pa_{v_2}\{\omega_1,\Omega_{B2}^\mu\}}{\pa_{v_2}\Omega_{B2}^\mu}\right)=\delta^{(3)}(x_{12}),
\eeq
which is evident in \eqref{s2p}. This shows that $\bT_f$ does not generically satisfy a closed algebra, as instructed from  \eqref{TTal}.  Noticeably, the algebra closes on non-expanding backgrounds, where $\Omega_\mu^B$ is constant in time, and thus $\{\omega_1,\omega_2\}=0$. Therefore, the phase space for expanding and non-expanding backgrounds are radically different. In the latter, the background area is a classical field while in the former the area expansion becomes a non-commutative field, which is key for quantization. Already at the classical level, we can understand this central result in the following way. On non-expanding backgrounds, the fluctuations of the area Poisson-commute with the area itself, as one can see from \eqref{omOm} setting $\Delta\Omega^\mu_{B21}=0$. Moreover, the area perturbation commutes with the spin-$2$ gravitons $X$ and $\bX$. Thus, the spin-$0$ and spin-$2$ sectors do not mix: the perturbative gravitons do not backreact on the geometry at this order of perturbation. Conversely, on expanding backgrounds, backreaction is already present, and the gravitons do not commute with the perturbative area. The constraint satisfying an algebra in turn forces the perturbative area to become non-commutative, realizing in formulas the idea that radiation on expanding backgrounds sources the expansion to change perturbatively, in a cascade effect.

Using the brackets among composite fields derived in App. \ref{A2}, with a lengthy computation, one proves the following properties
\beqn
\{\bC_f,\mu_2\}=\pa_{v_2}(f_2\mu_2)+\pa^2_{v_2}f_2,\qquad 
\{\bC_f,\Omega_{B2}^\mu\}=f_2\pa_{v_2}\Omega_{B2}^\mu,
\eeq
where we used the background constraint, $\pa^2_{v}\Omega_{B}^\mu-\mu\pa_{v}\Omega_{B}^\mu=C_{B}=0$. Performing similar manipulations, we gather
\beq
\{\bC_f,\pa_{v_2}X_2\}=\pa_{v_2}(f_2\pa_{v_2}X_2)\qquad \{\bC_f,\pa_{v_2}\bX_2\}=\pa_{v_2}(f_2\pa_{v_2}\bX_2),
\eeq
together with
\beq
\{\bC_f,\pa_{v_2}\omega_2\}=\pa_{v_2}(f_2\pa_{v_2}\omega_2)\qquad \{\bC_f,\pa^2_{v_2}\omega_2\}=\pa^2_{v_2}(f_2\pa_{v_2}\omega_2).
\eeq
Eventually, using these results, one computes
\beq
\{\bC_f,\bC_g\}=-\bC_{[f,g]}.
\eeq
This second way of deriving the constraint algebra is also an indirect proof of the exactness of the Poisson brackets derived in App. \ref{A2}. The derivation of the perturbative brackets for an expanding background is a novel result of this manuscript, and paves the way to the perturbative quantization of gravity on generic null hypersurfaces.

\subsubsection{Canonical Quantization: Non-expanding Background}

There are two compatible ways of setting the background area to be constant. One is to directly impose $\pa_v\Omega_B=0$, together with $\delta\Omega_B=0$, in the symplectic two form and re-derive the brackets. The second route is to directly impose $\pa_v\Omega^\mu_B=0$ in the brackets derived in the previous section. However, one has to recall that $\Omega_B^\mu$ is now independent of $\mu$, since the zeroth order constraint is automatically satisfied. Calling the background area element $\Omega_B$, these procedures yield the same result, that is,
\beqn
\{\omega_1,\omega_2\}&=&0\\
\{\omega_1,\Omega_{B2}\}&=&0\\
\{\omega_1,X_2\}&=&0\\
\{\omega_1,\bX_2\}&=&0\\
\{X_1,\bX_2\}&=&-\frac{\Theta(v_{21}) \delta^{(2)}(z_{12})}{2\sqrt{\Omega_{B1}\Omega_{B2}}}\\
\{\omega_1,\mu_2\}&=&\delta^{(3)}(x_{12}).
\eeq
Various comments are in order. First, we have that the area perturbation becomes a commutative field. This in turns implies that also $\bT_f$ satisfies a closed algebra,
\bee
\{\bT_f, \bT_g \} =  -\bT_{[f,g]}.
\eee
Secondly, the spin-$0$ and spin-$2$ contributions completely decouple: to this order, on non-expanding backgrounds, the spin-$2$ data are disentangled from the geometric perturbation of the background. Lastly, we note that the perturbative spin-$0$ sector behaves exactly like the finite spin-$0$ analysis of Section \ref{bg}. Moreover, the spin-$2$ data also couples together in a similar fashion, and indeed behave as a complex scalar field.

We can canonically quantize these Poisson brackets. The only non-zero commutators are
\beq
[\hat\omega_1,\hat \mu_2]= i\hbar \delta^{(3)}(x_{12})\qquad [\hat X_1,\hat\bX_2]=-\frac{i\hbar}{2 \sqrt{\Omega_{\B}{}_1\Omega_{\B}{}_2}}
\Theta(v_{21})\delta^{(2)}(z_{12}).
\eeq
Importantly, we assumed in this and previous sections that the background area element is required to be nowhere vanishing, and thus invertible, both classically and as an operator. This is natural when starting from a classical background. We will explore the consequences of releasing this assumption in Section \ref{mol}.
The spin-$0$ sector gives rise to a twisted $\beta\gamma$ CFT per null generator with
\beq\label{tt0}
\hat\omega=\frac{\hbar}{2\pi}\gamma, \qquad \hat\mu=\beta.
\eeq
It therefore contributes to the total central charge with $c_0=2$ per null generator. 

For the spin-$2$ sector, we again confine our attention to a single null generator, and following the same pattern we find the OPE
\beq
\pa_{v_1}\hat X_1\pa_{v_2}\hat\bX_2\sim\frac{\hbar}{4\pi\sqrt{\Omega_{\B}{}_1\Omega_{\B}{}_2}(v_{12}-i\epsilon)^2}.
\eeq
Given \eqref{T2pert}, we define the spin-$2$ quantum stress tensor
\beq\label{TX}
\hat \bT^X=\frac{4\pi}{\hbar}\Omega_B\pa_v \hat X\pa_v\hat\bX.
\eeq
We can then evaluate its OPE and gather
\beq\label{spin2br}
\hat \bT^X(v_1)\hat \bT^X(v_2)\sim \frac{c_2}{2(v_{12}-i\epsilon)^4}+\frac{2\, \hat \bT^X(v_2)}{(v_{12}-i\epsilon)^2}
+\frac{\pa_{v_2}\hat \bT^X(v_2)}{v_{12}-i\epsilon},
\eeq
with $c_2=2$. Therefore the spin-$2$ sector also contributes with a central charge $c_2=2$ per null generator. This is expected given that the perturbative graviton is a complex scalar field, and thus again the central charge counts the number of fields on each null ray.

Putting the spin-$0$ and spin-$2$ sectors together, we arrive at the main result. Defining the total quantum stress tensor
\beq
\hat \bC=\frac{2\pi\bC}{\hbar},
\eeq
we can compute the complete OPE and find
\beq
\hat\bC(v_1)\hat\bC(v_2)\sim \frac{c}{2(v_{12}-i\epsilon)^4}+\frac{2 \hat\bC(v_2)}{(v_{12}-i\epsilon)^2}
+\frac{\pa_{v_2}\hat\bC(v_2)}{v_{12}-i\epsilon},
\eeq
where
\beq
c=c_0+c_2+c^{\mathrm{mat}}=2+2+c^{\mathrm{mat}}.
\eeq
Here, $c^{\mathrm{mat}}$ is the matter contribution to the central charge, in case matter is added to the system.

Notice that one obtains the same result using
\beq
\hat \bT=\frac{2\pi}{\hbar}\bT,
\eeq
id est
\beq
\hat\bT(v_1)\hat\bT(v_2)\sim \frac{c}{2(v_{12}-i\epsilon)^4}+\frac{2\hat \bT(v_2)}{(v_{12}-i\epsilon)^2}
+\frac{\pa_{v_2}\hat\bT(v_2)}{v_{12}-i\epsilon}.
\eeq
This is due to the fact that on non-expanding backgrounds the area perturbation commutes, $\{\omega_1,\omega_2\}=0$. On the other hand, this ceases to hold on generic backgrounds, whereby the central charge can exclusively be computed using the constraint OPE. Similar to Section \ref{bg} and thanks to ultralocality, reintroducing the cut dependency gives
\beq
\hat\bC(x_1)\hat\bC(x_2)\sim \left(\frac{c\delta^{(2)}(0)}{2(v_{12}-i\epsilon)^4}+\frac{2 \hat\bC(x_2)}{(v_{12}-i\epsilon)^2}
+\frac{\pa_{v_2}\hat\bC(x_2)}{v_{12}-i\epsilon}\right)\delta^{(2)}(z_{12}).
\eeq
The addition of the spin-$2$ sector perturbatively does not interfere with ultralocality, and the central charge counting the degrees of freedom on each null generator. The only substantial difference is the number of such degrees of freedom, now given by $2+2+c^{\mathrm{mat}}$.

\subsubsection{Canonical Quantization: General Backgrounds} \label{metriplectic} 
The canonical quantization of the perturbative phase space on expanding backgrounds is more challenging, given the structure of the Poisson brackets. 
We leave its rigorous derivation for future works.
Although our phase space is infinite dimensional, we offer here some preliminary observations on quantization and the anomaly based on finite dimensional phase spaces.
Going from the phase space to the quantum algebra of observables always involves a departure from the Poisson commutators, except if the observables are linear in momenta, or if the observables are at most quadratic in the canonical variable and the polarization is chosen to be the Heisenberg one. Moreover, the quantum commutator depends on the choice of polarization. Here, we get some control of the anomaly of the constraint algebra, since the constraint \eqref{Tfirst} is at most cubic in the canonical fields.\footnote{We also need to take into account the fact that the target space for the spin-$2$ field is the homogeneous space $\mathrm{SL}(2,\mathbb{R})/ U(1)$.}

Given our symplectic data, to obtain the quantum commutator we can begin by defining the star product. If the symplectic structure is globally constant, the symplectic manifold is $\mathbb{R}^{2n}$ and we can use the Moyal star product \cite{Moyal_1949}. For any function $\phi,\psi\in \mathbb{R}^{2n}$ the latter is defined via
\beq
\phi\star \psi=\phi\psi+\sum _{n=1}^{\infty }\hbar ^{n}C_{n}(\phi,\psi),
\eeq
where the order-$n$ bidifferential can be written explicitly using Berezin's formula \cite{FABerezin_1974}. The first orders in $\hbar$ are
\beq
\phi\star \psi=\phi\psi+{\frac {i\hbar }{2}}\Pi ^{ij}(\partial _{i}\phi)(\partial _{j}\psi)-{\frac {\hbar ^{2}}{8}}\Pi ^{ij}\Pi ^{km}(\partial _{i}\partial _{k}\phi)(\partial _{j}\partial _{m}\psi)+\ldots.\label{fsg}
\eeq
Here, $i,j,\ldots$ are field space indices, and $\Pi$ is a Poisson bivector,
\beq
\Pi =\Pi ^{ij}\partial _{i}\wedge \partial _{j}.
\eeq
The matrix $\Pi^{ij}$ is nothing but the Poisson brackets on this symplectic manifold. 

There are two reasons why we need to generalize this formula. The first is that our symplectic manifold is not globally $\mathbb{R}^{2n}$ and so we should work only locally. One could locally try to find a change of coordinates to Darboux coordinates, as done by Fedosov \cite{fedosov1994simple, xu1998fedosov}, but this is not required, as we can directly use the Kontsevich quantization formula \cite{Kontsevich:1997vb}, which applies to an arbitrary Poisson manifold. The latter reads
\beq\label{Ksp}
\phi\star \psi=\phi\psi+\sum _{k=1}^{\infty }\hbar ^{k}B_{k}(\phi,\psi),
\eeq
for a new order-$k$ bidifferential $B_{k}(\phi,\psi)$. The first terms in the $\hbar$ expansion are given by
\beq\label{Kspex}
{\begin{aligned}\phi\star \psi&=\phi\psi+{\tfrac {i\hbar }{2}}\Pi ^{ij}\partial _{i}\phi\,\partial _{j}\psi-{\tfrac {\hbar ^{2}}{8}}\Pi ^{i_{1}j_{1}}\Pi ^{i_{2}j_{2}}\partial _{i_{1}}\,\partial _{i_{2}}\phi\partial _{j_{1}}\,\partial _{j_{2}}\psi\\&-{\tfrac {\hbar ^{2}}{12}}\Pi ^{i_{1}j_{1}}\partial _{j_{1}}\Pi ^{i_{2}j_{2}}(\partial _{i_{1}}\partial _{i_{2}}\phi\,\partial _{j_{2}}\psi-\partial _{i_{2}}\phi\,\partial _{i_{1}}\partial _{j_{2}}\psi)+{\mathcal {O}}(\hbar ^{3})\end{aligned}}
\eeq
which obviously reduces to \eqref{fsg} when the Poisson brackets are constant in phase space, that is when $\pa_i \Pi^{jk}=0$. With this star product, we can define the quantum commutator as\footnote{Strictly speaking, this cannot be applied directly in field theory, as the configuration space is infinite dimensional.}
\beq
[\phi,\psi]=\phi\star \psi-\psi\star \phi=i\hbar \{\phi,\psi\}+{\mathcal {O}}(\hbar ^{2}).
\eeq
This solves the first problem. The fact that the quantum commutator reproduces to first order the Poisson brackets is one of the pillars of the theory of deformation quantization.

The second issue is that we need to extend the star product to the space of microcausal functionals. The Poisson brackets $\Pi^{ij}$ must be upgraded to a K\"ahler structure
\beq
H^{ij}=G^{ij}+i\Pi^{ij},
\eeq
such that the singularity structure of $H^{ij}$ is only in the future. The precise formulation of this step is extensively explained using the wave front analysis in chapter $5$ of \cite{Rejzner:2016hdj}, see also references therein such as \cite{Brunetti:2001dx ,Hollands:2001nf, Fredenhagen:2012sb}. In simple terms, to understand the singularity structure in the quantization, we need a full symmetric and anti-symmetric structure in field space, while the Poisson brackets only inform the anti-symmetric part. Replacing $\Pi$ with $H$ in  \eqref{Ksp} gives a new product, called $\star_{H}$, that can be applied to microcausal functionals. The main point is that, on regular functionals, the two star products $\star$ and $\star_H$ are isomorphic: there exists a gauge transformation $\alpha_H$ such that
\beq
\phi\star_H \psi=\alpha_H\left((\alpha_H^{-1}\phi)\star(\alpha_H^{-1}\psi)\right).
\eeq
As is well-explained in \cite{Rejzner:2016hdj}, the physical interpretation of passing from $\star$ to $\star_H$ is the introduction of a notion of normal ordering since $\no{A}\no{B}= \no{A\star_H B}$. Therefore the $\star_H$ product is the algebraic version of Wick’s theorem. The normal ordering is equivalent to a choice of time, a choice of ground state and a choice of (almost) complex structure in holomorphic quantization, which we explore in the next section.

Eventually, we would like to explicitly evaluate \eqref{Kspex}. This is a long and intricate computation that we defer to future works. The expected outcome is
\beq
[\hat{\bC}_f, \hat{\bC}_g] = -\hat{\bC}_{[f,g]} - \frac{c}{12} \int_N\ve_{\cN}^{(0)} (f \pa_v^3 g- g\pa_v^3f),
\eeq
with 
\beq
\hat \bC_f=\frac{2\pi\bC_f}{\hbar}.
\eeq
The most important detail is to ensure that the central charge is non-zero and field independent, that is, it is still counting the degrees of freedom as in the non-expanding case.

In conclusion, this preliminary analysis shows that while the quantization on expanding backgrounds becomes more sophisticated, we expect that the essence remains the same as for non-expanding backgrounds: the constraints satisfy an algebra at the quantum level with a field-independent central extension counting the degrees of freedom. While we do not have a full proof of this result, we intend to return and perform this analysis in future works. Ultimately, we expect that the tools developed in the theory of Wick deformation quantization will be of central relevance \cite{dolgushev2001wick}. In particular, these techniques can be applied directly to the non-perturbative phase space.

\subsection{Asymptotic Infinity}\label{sec: asymptinfty}

In this section we study the consequences of our results for asymptotic null infinity. 
Following e.g. the conventions of \cite{Freidel:2021fxf}, the metric of an asymptotically flat spacetime in Bondi coordinates is given to second order in the radial expansion by 
\be 
\rd s^2 = - 2 e^\beta  \rd u \left( \rd r  +\left(\frac12 - \frac{m_{\mathsf{B}}}{r}\right) \rd u \right) 
+ r^2 \gamma_{AB} \left( \rd \sigma^A +\frac1{ r^2} \Pi^A \rd u \right) 
\left(\rd \sigma^B +\frac1{ r^2}\Pi^B \rd u \right),
\ee 
where the fields admit the following asymptotic expansion
\bea 
e^\beta &=& 1- \frac1{8 r^2} C_{AB}C^{AB} +\ldots \\
\Pi^A &=& \frac12 D_B C^{BA} +\ldots \\
\gamma_{AB} &=& q_{AB} + \frac1r C_{AB} + \frac{1}{4 r^2}C_{CD}C^{CD} q_{AB} + \ldots.
\eea  
Here $q_{AB}$ is the time independent round sphere metric and the metric expansion implies that $\det(\gamma)=\det(q)$. This metric  converges toward the flat space metric at $\iota^+$. This means, once an adapted supertranslation frame is chosen, that 
\be 
\lim_{u\to +\infty}m_{\mathsf{B}}(u,\sigma)= 0, \qquad 
\lim_{u\to +\infty}C_{AB}(u,\sigma)= 0.
\ee
Note that the limit to $\iota_0$ of $m_{\mathsf{B}}$ and $C_{AB}$ is non-vanishing and respectively given by the ADM mass and the memory observable. As customary at asymptotic null infinity, we have denoted the null time by $u$, such that the boundary coordinates are $(u,\sigma^A)$. We also introduce the news tensor $N_{AB}=\pa_u C_{AB}$.

At asymptotic null infinity the time evolution of the energy is controlled by the Bondi mass loss formula 
\be \label{bm}
\pa_u m_{\mathsf{B}}= \frac14 D_AD_B N^{AB} - T^{\mathsf{H}}, \qquad  T^{\mathsf{H}}:= \frac18 N_{AB}N^{AB} + 8\pi G T_{uu}^{\mathrm{mat}},
\ee 
where $T^{\mathsf{H}}$ represents the matter and spin-2 hard degrees of freedom. 
A convenient way to write this equation is to introduce the Moreschi mass $m_{\mathsf{M}}:= m_{\mathsf{B}}-\frac14 D_AD_B C^{AB} $ for which \eqref{bm} simply reads $ \pa_u m_{\mathsf{M}}= - T^{\mathsf{H}}$.

In \cite{Kapec:2016aqd}, the authors have established the important result that the renormalized area of the asymptotic sphere is directly related to the time integral of the Moreschi mass. We briefly re-derive their result here locally on the cut. Consider 
the hypersurface $r=r(u_0)$ satisfying 
$ \pa_u r^2 =  2m_{\mathsf{B}} - r$. This is a null hypersurface approaching asymptotic infinity in the limit $u_0\to \infty$.
Integrating this equation means that 
\be 
r^2(u) = \frac{(u-u_0)^2}{4} + 2 \pa_u^{-1} m_{\mathsf{B}}(u) +\ldots, \qquad \pa_u^{-1} m_{\mathsf{B}}(u):= -
\int^{+\infty}_u \rd u' m_{\mathsf{B}}. 
\ee 
The induced  metric on the cuts $C_T=\{u= T(\sigma)\}$ is
\be 
Q_{AB} = r^2(T) \gamma_{CD}\left(\delta^C_A +\frac1{r^2} \Pi^C \pa_A T\right)\left(\delta^D_B +\frac1{r^2} \Pi^D \pa_B T\right),
\ee
where $\Pi^A=\frac12 D_B C^{BA}$ to leading asymptotic order. Consequently
\be 
\Omega(T) = \sqrt{\det{Q}} = \sqrt{q}\left(r^2(T) +   \Pi^A D_A T \right) + \ldots
\ee 
We can renormalize this area using the flat space contribution, $\Omega_R =\Omega-\frac{(u-u_0)^2}{4}$, thus obtaining
\bea\label{aren}
\Omega_R(T)=\left[\pa_u^{-1}\left( 2 m_{\mathsf{B}} +    J^A  D_A T  \right)\right](T),
\eea
where we denoted $J^A= \pa_u\Pi^A$ the current and we denote $[\pa_u^{-1}O](T) := \int^{T}_{+\infty} O$. If $\lim_{u\to +\infty} u m_{\mathsf{B}}=\lim_{u\to +\infty} u N_{AB}=0$, we can evaluate 
\beq
&\pa_u^{-1} m_{\mathsf{B}}(T)=  
\int_T^{+\infty} (u-T) \pa_u m_{\mathsf{B}}&\\ 
&\pa_u^{-1}(J^A  D_A T)= \int_T^{+\infty} 
J^AD_A(u-T)
= -\int_T^{+\infty} 
(u-T) D_AJ^A  + 
D_A M^A(T),&
\eeq
where we introduced a ``memory term'' $M^A(T) := \int_T^{+\infty} 
J^A(u-T) \rd u$.
This term vanishes when integrated over the cut, and so it does not contribute to the total area derived in \cite{Kapec:2016aqd}.

Putting these results in \eqref{aren}, and using \eqref{bm}, we derive the renormalized area element
\bea\label{or}
\Omega_R(T)=
\int_T^{+\infty}   (u-T)  (2\pa_u m_{\mathsf{B}} -   D_AJ^A) 
+ D_AM^A(T) = 
-2 \int_T^{+\infty}  (u-T) T^{\mathsf{H}}+ D_AM^A(T). 
\eea
Notice that we previously derived that the area is the boost Hamiltonian, and we are here seeing a confirmation of this result from asymptotic infinity, since we have found, following \cite{Kapec:2016aqd}, that the renormalized area is the integral of the boost aspect $(u-T) T^{\mathsf{H}}$, plus a total derivative term. Another remark is that the renormalized area is entirely dictated by the graviton distribution at infinity. In Section \ref{mol}, we will introduce the concept of embadon, as a quantum area bit on the cut. Here, the asymptotic analysis shows that the embadon attains its maximal value in flat space, and then decreases due to the presence of gravitons. A suggestive picture is that the embadon decreases over time because it emits gravitons and thus loses degrees of freedom.

The quantization of the asymptotic phase space at null infinity has been achieved in \cite{Ashtekar1981,Dappiaggi:2005ci,  Ashtekar:2014zsa, Strominger:2017zoo, Ashtekar:2018lor, Hollands:2019ajl, Laddha:2020kvp, Prabhu:2022zcr}, see also \cite{Bousso:1999cb}. The asymptotic time is defined to be the Bondi time and the $2$ independent components of  the news operator are the time derivative of a complex scalar. Therefore, calling $N_{zz}=N$ and $N_{\bz\bz}=\bN$, the OPE is
\be 
N(x_1) \bN(x_2) \sim \frac{4\delta^{(2)}(\sigma_{12})}{(u_{12}-i\epsilon)^2} .
\ee 
This result appeared in many works on asymptotic infinity, see e.g.  \cite{Hollands:2019ajl, Laddha:2020kvp, Prabhu:2022zcr,kay1991theorems, dappiaggi2017hadamard, Freidel:2022skz}. Assuming no matter is present, from \eqref{bm} we can then compute the OPE
\be 
T^{\mathsf{H}}(x_1)T^{\mathsf{H}}(x_2) \sim \left(\frac{\delta^{(2)}(0)}{(u_{12}-i\epsilon)^4}+\frac{2T^{\mathsf{H}}(x_2)}{(u_{12}-i\epsilon)^2}+\frac{\pa_{u_2}T^{\mathsf{H}}(x_2)}{u_{12}-i\epsilon}\right)\delta^{(2)}(\sigma_{12}).
\ee 
This is the analog at null infinity of \eqref{spin2br}, and shows that there is central charge $c_2=2$ on each null generator for the radiative sector. This  entails an infinite central charge due to the infinitely many null generators. 

Using the Bondi mass loss formula, this is also the OPE of the time derivative of the Moreschi mass operator
\beq
\pa_{u_1}m_{\mathsf{M}}(x_1)\pa_{u_2}m_{\mathsf{M}}(x_2) \sim \left(\frac{\delta^{(2)}(0)}{(u_{12}-i\epsilon)^4}-\frac{2\pa_{u_2}m_{\mathsf{M}}(x_2)}{(u_{12}-i\epsilon)^2}-\frac{\pa^2_{u_2}m_{\mathsf{M}}(x_2)}{u_{12}-i\epsilon}\right)\delta^{(2)}(\sigma_{12}).
\eeq
The asymptotic mass has therefore infinite fluctuation,\footnote{We acknowledge private discussions with Suvrat Raju on this point.} controlled by
\beq
m_{\mathsf{M}}(x_1)m_{\mathsf{M}}(x_2) \sim -\frac{\delta^{(2)}(0)\delta^{(2)}(\sigma_{12})}{6(u_{12}-i\epsilon)^2} +\ldots
\eeq
This result is expected to have relevant repercussions in celestial holography. Indeed, although we have not yet computed the area fluctuation, it is now a well-posed problem. Note that in the limit $u_i \to \infty$ this divergence can be argue to cancel \cite{Prabhu:2022zcr} and we recover the fact that the ADM mass is a well defined operator with finite fluctuation. 
One still has even in this limit non-zero fluctuations due to the neglected terms in the ellipsis. Understanding the nature and scale of these fluctuations is important for a deeper understanding of the black hole evaporation process.

This result establishing infinite fluctuation of the Bondi mass was anticipated by Wall and Bousso in \cite{Bousso:2017xyo}, where the authors argued that the Bondi mass does not exist as an operator. What our result shows is that the infinite fluctuation of the Bondi mass is expected from its interpretation as a modular Hamiltonian. Indeed, it is well established that due to the type III nature of QFT, the modular Hamiltonian  has infinite fluctuation \cite{Witten:2021unn}. Here we understand this fact as being the consequence of an infinite central charge.

\section{Quantum Time}\label{qutime}

In this section we discuss the appearance of time at the quantum level due to the central charge in the algebra of the quantum constraints. 

\subsection{The Problem of Time\ldots}\label{PofT}

One of the central problems in quantum gravity is the problem of time \cite{isham1993canonical, anderson2012problem, Carlip:2023daf}. One usually assumes that at the quantum level the constraints are satisfied without anomaly. Therefore, implementing the time evolution constraints means that no fundamental notion of time exists: physical observables must be time-independent, while states must be invariant under time reparametrizations. The first part of the problem of time, about observables, is by now essentially resolved \cite{rovelli1991time, Giddings:2005id, Gambini:2006yj, Dittrich:2006ee, hohn2021trinity}: one needs to construct physical observables as relational observables. Such a procedure requires decomposing the system into a clock and a physical subsystem. The physical subsystem elements then evolve relative to the clock subsystem. As shown in \cite{Ciambelli:2023mir},  a clock subsystem is readily available for gravitational physics along a null surface in terms of spin-$0$ data where a physical time $V$ was defined satisfying $\pa^2_v V = \mu \pa_vV$; this can be viewed as a feature of the extended phase space \cite{Klinger:2023qna,Klinger:2023auu,AliAhmad:2024wja}.
The clock time is then a physical observable. As such, it is necessarily subject to quantum fluctuations that can smooth out some of the UV divergences of QFT \cite{Witten:2021unn ,Chandrasekaran:2022cip,Jensen:2023yxy,Klinger:2023auu,Faulkner:2024gst,Kudler-Flam:2023hkl, DeVuyst:2024pop}.

The second part of the problem of time, i.e., the time reparametrization invariance of physical states, is a problem which still needs to be solved. It is a problem which lies at the core of preventing non-perturbative approaches to quantum gravity from agreeing with QFT techniques and Fock-like quantization.\footnote{The incompatibility between quantum gravity states appearing in loop quantum gravity (LQG) and the usual Fock states has been the subject of many debates  \cite{Nicolai:2006id, Thiemann:2006cf}.
For example, Thiemann emphasizes in \cite{Thiemann:2006cf} ``in LQG, we cannot use perturbation theory, Fock spaces, background metrics, etc... This is not the fault of LQG. It will be a common feature of all quantum gravity theories which preserve background independence.''
This tension also manifests itself in the discussion around the Kodama state, which provides a non-perturbative solution of the Wheeler-de-Witt constraints
\cite{Witten:2003mb,Freidel:2003pu, Alexander:2022ocp}. 
See \cite{Thiemann:2024tmv} for a recent discussion on this issue in quantum gravity.} 
Given the ongoing and dynamic nature of research in quantum gravity, and the central importance of the problem of time as a stumbling block to our understanding, let us delve further into this point. 

There are three related cornerstones in building a relativistic quantum field theory. The first is the choice of Fock ground state, which requires the choice of background time to decompose fields into positive and negative frequency components. The ground state is defined as the state annihilated by the negative energy field modes. The second foundational property is that the Fock ground state and any high energy state are thermally entangled across subregions \cite{Unruh:1983ms}.
The third one is that the parity reversing operator $P$ is quantized as a unitary operator, while the time reversal operator is quantized as an anti-unitary operator. Each of these features points towards a quantization scheme that breaks diffeomorphism invariance in some way.

Let us begin with the choice of ground state, reviewing some familiar features of free field theories: in order to quantize a physical system represented by a symplectic potential, one has to select a polarization. In field theory, the only polarizations available are complex polarizations, which require, in agreement with the geometric quantization program,  equipping the classical phase space with a complex structure \cite{Bowick:1987wg, woodhouse1992geometric}. 

The field space complex structure $I$, satisfying $I^2=-1$, that defines the Fock ground state in  a quantum field theory is given by 
$I(\phi)= iP_+\phi-iP_-\phi$ where $P_\pm$ are  projectors onto positive and negative frequency modes of $\phi$. Given a null time $v\in \mathbb{R}$, the positive frequency projectors are given by the non-local time integral 
\be 
(P_{+}\phi)(v) =\int_{-\infty}^{+\infty}\frac{\rd u}{2\pi i } \frac{\phi(u)}{v-u-i\epsilon}, \qquad P_- = P_+^\dagger.
\ee 
Such a positive frequency projector is not invariant under diffeomorphisms. Indeed, denoting $\hat C_f=\int_{-\infty}^{+\infty} \rd v\, f(v) \hat C$ the infinitesimal generator of the diffeomorphism $f\pa_v$ and $F_f(v)$  the corresponding finite diffeomorphism, a weight-$0$ scalar field transforms as
\beq\label{field}
e^{i\hat C_f} \phi e^{-i \hat C_f} = \phi \circ F_f.
\eeq
Therefore, one gets 
\be 
[e^{i \hat C_f} P_+ e^{-i\hat C_f}] \phi(v)  = 
\int_{-\infty}^{+\infty}\frac{\rd u}{2\pi i} \frac{\pa_u F_f(u)\, \phi(u)}{F_f(v)-F_f(u)-i\epsilon} \neq P_+\phi(v),
\ee 
where we have assumed that the domain of $F_f(v)$ is the same as the domain of $v$, that is, $\mathbb{R}$.

This non-trivial transformation property of the projector shows that a choice of ground state is not a notion that is invariant under diffeomorphisms. Indeed, one of the hallmarks of QFT states is that diffeomorphisms changing the time integration domain are associated with thermal behaviour \cite{PhysRevD.7.2850, Unruh:1976db}.
For instance, consider the transformation $F_f(v)= e^{\kappa v}$. It converts a translation $\pa_v$ in $(0,+\infty)$ into a  boost vector field $\kappa F_f\pa_{F_f}$ on $\mathbb{R}$. It also maps the ground state  at the quantum level into a thermal state with temperature $\kappa/2\pi$. The fact that after the map $F_f$ the state is thermal follows from the fact that $F_f(v-2\pi i/ \kappa)=F_f(v)$, which implies the KMS condition, i.e., thermality.

We have seen that the choice of time determines the ground state. Remarkably, the converse is also true. This is the thermal time hypothesis formulated by Connes and Rovelli \cite{Connes:1994hv}, stating that a choice of state is equivalent to a choice of time, and formulated using the Tomita-Takesaki theory \cite{Takesaki:1970aki}. 
This result emphasizes that in QFT the ground state depends on the chosen time.

For the arguments presented here, it is instructive to recall the behaviour under time reparametrization of the two-point function of a free complex scalar field $\phi$.
Given a Fock ground state $|0\rangle$ in some given time $v$, we denote by  $|F_f\rangle_{{}_0}:=e^{-i\hat C_f} |0\rangle$ the ground state associated with the time $F_f(v)$.
Given \eqref{field}, one has
\be 
{}_{{}_0}\langle F_f| \pa_v\hat\phi_1 \pa_v\hat{\bar{\phi}}_2|F_f \rangle_{{}_0}= 
\langle 0| e^{i\hat C_f} \pa_v\hat\phi_1 \pa_v\hat{\bar{\phi}}_2 e^{-i\hat C_f}|0\rangle= \frac{\pa_vF_f(v_1) \pa_vF_f(v_2)}{(F_f(v_1)-F_f(v_2)-i\epsilon)^2 } .
\ee 
That is, $\pa_v \hat\phi$ behaves as a primary field of weight one. The creation operator $\hat a^\dagger_\omega$ associated with the ground state $|0\rangle$ is related to the creation operator $\hat a^\dagger{}^f_{\omega'}$ associated with the ground state $|F_f \rangle_{{}_0}$ via the Bogoliubov coefficients 
\be 
\beta_{\omega \omega'}= \int_{-\infty}^{+\infty}\rd v\, e^{i\omega v } e^{i \omega' F_f(v)}.
\ee 
If $||\beta||^2< \infty $, then the transformation is inner \cite{Birrell:1982ix}. We emphasize that there is nothing special about the state that we called $|0\rangle$: it is on the same footing as any of the $|F_f\rangle_{{}_0}$. Given a choice of such a state, we can discuss physics in relation \cite{Aharonov:1984zz, Rovelli:1990pi, Kitaev:2003zj, Bartlett:2006tzx, Girelli:2007xn, Palmer:2013zza, Giacomini:2017zju, Vanrietvelde:2018pgb, Hoehn:2019fsy, delaHamette:2020dyi, Giacomini:2021gei, delaHamette:2021oex, AliAhmad:2021adn, Carrozza:2021gju, Goeller:2022rsx,AliAhmad:2024wja} to that state. 

Denoting $H_{F_f}(v_1,v_2):={}_{{}_0}\langle F_f| \pa_v\hat\phi_1 \pa_v\hat{\phi}_2|F_f\rangle_{{}_0}$ the correlation function in the ground state $|F_f\rangle_{{}_0}$, one of the important properties of such a family of ground states is that the difference between two expectation values is free of divergences. Remarkably, one finds that \cite{davies1976energy,birrell1984quantum, Bianchi:2014qua}
\be \label{limS}
\lim_{v_1\to v_2}( H_{F_f}(v_1,v_2) - H_{0}(v_1,v_2)) 
= \frac16 \{F_f; v_2\},
\ee
where we recall that $\{F_f;v \} = \pa_v\left(\frac{\pa_v^2F_f}{\pa_vF_f}\right)- \frac12\left(\frac{\pa_v^2F_f}{\pa_vF_f}\right)^2 $ is the Schwarzian derivative.

The appearance of a central charge in the constraint algebra allows us to propose a resolution of the fundamental tension between the ground state definition in QFT, which depends on the time chosen, and the sought-for background independence in quantum gravity.
Indeed, the usual perspective is that background independence implies the absence of a physical notion of time \cite{Rovelli:2009ee}, and choosing   a preferred ground state $|0\rangle$ leads in one way or another to a violation of background independence. These statements are based on the usual idea that the diffeomorphism constraints can be applied either classically or quantum mechanically. 
As we have seen, the main point now is that the constraint algebra is \emph{anomalous} at the quantum level. A central charge is produced in the quantization scheme due to the presence of normal ordering necessary to define the constraint operator 
\be 
[\hat{C}_f, \hat{C}_g] = -\hat{C}_{[f,g]} - \frac{c}{12} \int_{\cN}\ve_{\cN}^{(0)} (f \pa_v^3 g- g\pa_v^3f),
\ee 
which implies, calling $F_f(v)=v'(v)$ to make contact with standard CFT notation,
\beq\label{lgdiffonconstr}
\hat C'(v')=\left(\pa_v v'\right)^{-2}\Big(\hat C(v)-\frac{c}{12}\{v';v\}\Big).
\eeq

At the classical level $c_{\mathrm{class}}=0$; in perturbative QFT on the other hand we have that 
$c_{\mathrm{QFT}}=\infty$.
We postulate that in quantum gravity, $c_{\mathrm{QG}}=c N$  is finite and is a measure of its self-regulated nature. In $c_{\mathrm{QG}}$, $c$ is a count of the number of fundamental degrees of freedom on each null generator.
Thus the presence of a central charge in the constraint algebra radically changes the discussion around the time problem, and as we will show, provides a simple and natural resolution of it.

First, it is of crucial importance that the anomaly we find in the constraint algebra is associated to a \emph{central charge} and not a field-dependent cocycle, which would lead to an impasse in the quantization.\footnote{A field-dependent cocycle is a deformation of the algebra $[\hat{C}_f, \hat{C}_g]= 
-\hat{C}_{[f,g]} + K_{(f,g)}$ where $K$ is not central, that is, $[K_{(f,g)}, \hat O]\neq 0$, see \cite{Barnich:2010eb,Chandrasekaran:2020wwn,Freidel:2021cjp}.}
The fact that the algebra deformation is central means that the notion of physical operators is unchanged since we can still define gauge invariant operators to be time-independent. In other words, $\hat O$ is physical if and only if $[\hat{C}_f,\hat O]=0$. This is consistent since, if the algebra deformation is central, the commutator algebra is unmodified
\be 
[\hat{C}_f,[\hat{C}_g,\hat O]]- 
[\hat{C}_g,[\hat{C}_f,\hat O]] = -[\hat{C}_{[f,g]},\hat O].
\ee
Therefore, the presence of a central charge does not modify the notion of gauge invariant observables. In particular, it does not change the `number' of degrees of freedom measured by those observables.

On the other hand, the central charge modifies discussion about states. It is no longer possible  to impose at the quantum level that $\hat{C}_f |\psi\rangle =0$ for all $f$. We now have to select a projective representation of $\mathrm{Diff}(\mathbb{R})$ associated with the given central charge. In particular one finds that 
\be \label{Bottrep}
e^{-i \hat {C}_f}|G_g \rangle_{{}_0} = e^{i cN  B(F_f,G_g) } | F_f \circ G_g\rangle_{{}_0},
\ee 
where we recall $|G_g \rangle_{{}_0}=e^{-i\hat C_g}|0 \rangle$, $|F_f\circ G_g \rangle_{{}_0}=e^{-i\hat C_{[f,g]}}|0 \rangle$, and
\be \label{Bottcocycle}
B(F_f,G_g)= -\frac1{48\pi} \int \rd v \ln[ (\pa_vF_f) \circ G_g] \pa_v \ln[\pa_v G_g]
\ee
is the Bott-Thurston cocycle \cite{bott1977characteristic,Oblak:2017ect}.

As already mentioned, we stress that the anomalous transformation is the rescaling of $\ell$, which is used in the primed phase space to render $\ell$ a background structure, $\delta\ell=0$. On the other hand, diffeomorphisms of $\cN$ are not anomalous, consistent with the absence of a gravitational anomaly \cite{Bilal:2008qx, Alvarez-Gaume:1983ihn}. The primed diffeomorphisms used in our construction are a combination of diffeomorphisms and rescalings, and therefore are anomalous. This is analogous to the Weyl anomaly in ordinary CFTs, where the Weyl transformation is a global symmetry used to compensate conformal isometries, in order to preserve the background metric.

\subsection{\ldots and its Resolution}\label{ctime}

In the previous section we have seen that the quantization depends on the choice of time. Furthermore, we have seen that the central charge induces a {\it projective} representation of the primed diffeomorphisms. In Section \ref{conftime}, we have discussed choices of dressing times and argued that conformal time is particularly significant in that it is the unique choice for which $\mu_c$ remains invariant under conformal transformations. Correspondingly, for time reparametrizations on ${\cal N}$, there is an $\mathfrak{sl}(2,\RR)$ subalgebra 
that arises as a `global' symmetry in conformal time. 

We will be interested in constructing a suitable Hilbert space, with this $\mathfrak{sl}(2,\RR)$ as a guide. Since we regard this as a global symmetry, physical Hilbert space states should be associated with representations of $\mathfrak{sl}(2,\RR)$. Note that this is not a statement that one would usually make, given that we regard diffeomorphisms as gauge symmetries; however, we must take into account the fact that the constraints projectively represent the action of {\it primed} diffeomorphisms.  Suppose we have a state $|\emptyset\rangle$, and consider the expectation value of the centrally extended OPE in this state. One immediately sees that it is not possible to set $\hat C_f|\emptyset\rangle=0$ for all $f$. Instead, what one can do is split the constraints into two sets, one which annihilates $|\emptyset\rangle$, and one which does not (instead creating physical finite norm states). As discussed in Section \ref{metriplectic}, one should regard this as specifying that the theory must be quantized in a holomorphic polarization, and should be interpreted as the introduction of a particular notion of projectors $P_\pm$ analogous to those discussed in Section \ref{PofT}. What is most important though is the fact that such a choice of projectors is associated with particular $\mathfrak{sl}(2,\RR)$ representations. We will propose that there is a  quantization in which there is an $\mathfrak{sl}(2,\RR)$-invariant state. 

The identification of the $\mathfrak{sl}(2,\RR)$-invariant state allows us to clarify the role of the Raychaudhuri constraint. It is useful to consider the familiar case of the quantization of 2d conformal field theories on a circle as a warm-up exercise: in that case, there is a convenient choice of basis $f_n\sim e^{inv}$, and the modes of the stress tensor are the discrete $L_n, n\in \ZZ$ where $L_{-1},L_0,L_{+1}$ generate an $\mathfrak{sl}(2,\RR)$ subalgebra. The $\mathfrak{sl}(2,\RR)$-invariant vacuum state satisfies $L_{n}|\emptyset\rangle=0$ for $n=-1,0,1,2,...$ whereas $L_{-n}|\emptyset\rangle$ for $n=2,3,...$ are (generally) physical. The significant structure here is the splitting of the modes into two subsets, one which annihilates the vacuum and one which does not. The algebra of the $L_{n}$'s possesses a $*$-involution under which $L_{n}^*=L_{-n}$, and it is its existence which allows us to properly invoke the constraint in the quantum theory in the presence of a central charge. That is, the vacuum expectation value of the constraint vanishes.

The construction in our case proceeds analogously. Since we are quantizing on the real line instead of the circle, some details are different. As is familiar, there are many available representations of $SL(2,\RR)$, and as a non-commutative algebra, it is convenient to interpret a unitary (projective)  representation on a Hilbert space ${\cal H}$ as defining a map  $\pi:\mathsf{Diff}({\cal N})\to U({\cal H})$.

Given the discussion in Section \ref{conftime}, we consider the real line parameterized in conformal time, here renamed $v_c$, which is a caustic avoiding time, $v_c\in(-\infty,+\infty)$. We then write as before
\beq
\hat C_f =\int^\infty_{-\infty} \rd v_c f(v_c)\hat C(v_c).
\eeq
We will interpret $\hat C_f$ as a Hermitian operator for all vector fields generating time reparameterizations $\ell=f(v_c)\pa_{v_c}$.
We are interested here in introducing a suitable basis to express the functions $f$. One might have expected that this formula implies that $\hat C_f$ is a tensor.  This is not quite true, as is familiar in any 2d CFT. Recalling  \eqref{lgdiffonconstr}, and considering a diffeomorphism $v_c\to v'(v_c)$ generated by $f$; we have
\beqn
\hat C'_{f'}=\int\rd v' f'(v')\hat C'(v')=\hat C_{f}-\frac{c}{12}
\int^\infty_{-\infty}\rd v_c f(v_c)\{v';v_c\},
\eeqn
where we used that $f$ is the component of a vector field, $f'(v')=(\pa_{v_c}v') f(v_c)$.

This will mean that operators $e^{-i\hat C_f}$ will furnish a projective representation of time reparameterizations, which contain inside  a representation of $SL(2,\RR)$ generated by the $\hat C_f$ with $f\in \mathfrak{sl}(2,\RR)$.
That is, if we restrict attention to $SL(2,\RR)$, the Schwarzian vanishes and thus $\hat C_f$ is a tensor.  

Now we consider the corresponding representation space ${\cal H}$. It is useful to consider a basis that diagonalizes $v_c\pa_{v_c}$.\footnote{Another choice of basis would be the Fourier basis $e^{-i\omega v_c}$ for which $i\pa_{v_c}$ is diagonal and Hermitian if $\omega\in\RR$. We use the following basis instead, because then the $\mathfrak{sl}(2,\RR)$ generators correspond to $f(v_c)=(1,v_c,v_c^2)$ rather than $e^{inv_c}$ for $n=-1,0,1$. These two bases can be thought of as being related via a diffeomorphism with non-zero Schwarzian.} The eigenfunctions $\phi_\Delta(v_c)\in C^{\infty}(\RR)$ are of the form $\phi_\Delta(v_c)\sim v_c^\Delta$ with eigenvalue $\Delta$. The spectrum of $\Delta$ is as yet undetermined. The most important assumption is that for each $\phi_\Delta$, there is a corresponding $\phi_{\Delta_*}$ where $\Delta_*$ is the dual value. While other representations constructed by acting with primary operators will be considered shortly, the reader may anticipate that given that we are talking  here about modes of $\hat C(v)$, we will be interested  for now in a particular representation which includes all such modes, in which $\Delta$ is real. In the present situation, one has $\Delta_*=2-\Delta$.  It is then convenient to relabel $\Delta=1+\lambda$, in terms of which $\lambda_*=-\lambda$ and thus we will assume that the spectrum is real and symmetric about the origin in $\lambda$.
We write the formal expression\footnote{Here we are suppressing issues of convergence. In particular, this expression is formal because it is not a tempered distribution, that is, $f$ is not rapidly decreasing. One way to resolve this is to include a  regulator factor $e^{-\epsilon |v_c|}$. This is the same mechanism at the origin of the $v_{12}-i\epsilon$ factors in correlators.}
\beq\label{defClambda}
 \hat C_\lambda=\int_{-\infty}^{\infty} \rd v_c\, v_c^{\lambda+1}\hat C(v_c).
 \eeq
In this basis the $*$-involution corresponds to inversion  $v_c\mapsto -1/v_c$. This implies 
\beq
\hat C_\lambda^*=\hat C_{-\lambda},
\eeq
which follows from previous formulas, noting that the inversion is in fact an element of $SL(2,\RR)$.

The $\mathfrak{sl}(2,\RR)$-invariant vacuum then satisfies
\beq
\hat C_\lambda|\emptyset\rangle=0,\qquad \forall \lambda \geq -1.
\eeq
There is a close analogy here with ordinary 2d CFT. What we are describing is the vacuum module of the chiral conformal symmetry. The vacuum state $|\emptyset\rangle$ should not be confused with a Fock ground state $|0\rangle$. Indeed, this vacuum has the special feature of being the unique $\mathfrak{sl}(2,\RR)$-invariant state in the module, corresponding to the identity operator of the CFT. 
The physical states then are generated by acting on the vacuum with $\hat C_{-\lambda}$ for $\lambda > 1$, which correspond to  conformal descendants of the identity. As we described above, it is useful to introduce the `orientation states' associated with diffeomorphism group elements
\beq
|F_f\rangle = e^{-i\hat C_f}|\emptyset\rangle.
\eeq
This is an overcomplete basis of states which satisfy \eqref{Bottrep},
\beq\label{orstoverlap}
\langle F_f | G_g\rangle =  e^{i cN  B(F^{-1}_f,G_g) }\langle\emptyset |\emptyset\rangle.
\eeq

It is interesting to consider what these states correspond to physically. Given that they are defined by acting with the unitary operator $e^{-i\hat C_f}$ on the $\mathfrak{sl}(2,\RR)$-invariant state, it is natural to interpret them as a ground state from the perspective of a different time. Thus in this construction, all such states are in fact physical and related to the $\mathfrak{sl}(2,\RR)$-invariant state by symmetry. The vacuum module constructs all such states. This is in contrast to what happens in a non-conformal quantum field theory where the Fock ground state is associated with a choice of time and consequently a choice of such a state in which to quantize the theory. Then, the problem of time can be viewed from this perspective as the fact that diffeomorphisms (here just time reparameterizations) act non-trivially on such states. In fact, as discussed for example in \cite{AliAhmad:2024wja}, the states $|F_f\rangle$ correspond to quantum reference frames, see \cite{Aharonov:1984zz, Rovelli:1990pi, Kitaev:2003zj, Bartlett:2006tzx, Girelli:2007xn, Palmer:2013zza, Giacomini:2017zju, Vanrietvelde:2018pgb, Hoehn:2019fsy, delaHamette:2020dyi, Giacomini:2021gei, delaHamette:2021oex, AliAhmad:2021adn, Carrozza:2021gju, Goeller:2022rsx}.

Now suppose that we considered an excited state, which we think of as obtained by acting with a field operator on the vacuum. For instance, we can think of the graviton excitation $\hat{\tilde\sigma}$, although the precise details are not relevant here. As discussed in Section \ref{conftime} (see also \cite{Ciambelli:2023mir}) we recall that in going to a dressing time such as conformal time, operators are dressed to the time reparameterization diffeomorphisms, so they are invariant under the constraints,
\beq
e^{i\hat C_f}\hat{\tilde\sigma}e^{-i\hat C_f}=\hat{\tilde\sigma},
\eeq
which is the finite version of \eqref{vestito}.
Therefore, we can always think of the corresponding excited states as arising from acting on the $\mathfrak{sl}(2,\RR)$-invariant vacuum, $|\tilde\sigma,\emptyset\rangle = \hat{\tilde\sigma}|\emptyset\rangle$, 
such that
\beq
e^{i\hat C_f}|\tilde\sigma,\emptyset\rangle = \hat{\tilde\sigma}|F_f\rangle = |\tilde\sigma,F_f\rangle.
\eeq
This  means that a dressed excitation can be covariantly defined on any state in the vacuum module. Stated differently, the dressed graviton can be expressed in any time frame, and the relation between the states in different frames is given as usual in terms of Bogoliubov  coefficients. This again is a reflection of the fact that the notion of quantum reference frames are provided by the dressing here. The novelty of the current construction is that there is in fact a unique vacuum $|\emptyset\rangle$ from which all other states $|F_f\rangle$ in the vacuum module are built.  

In our discussion, we have implicitly considered the central charge as finite. Consequently, we have found that all orientation states $|F_f\rangle$ are on the same footing, and in fact we can consider arbitrary superpositions of them. This stands in stark contrast to the usual assumptions in quantum field theory, where one makes a choice of ground state $|F_f\rangle_{{}_0}$ (and a corresponding time) in which to quantize the theory. These ground states are regarded as defining a superselection sector, all such states being orthogonal.\footnote{If we associate the central charge with a notion of energy (such as Casimir energy), an infinite central charge implies that transitioning from one state to another within the vacuum module requires an infinite amount of energy. This results in a fixed background in the classical limit. Conversely, at the quantum level, a finite central charge means the energy gap between these backgrounds is finite, allowing for their coexistence.} This usual approach leads to a violation of covariance or background independence when applied to the usual field theoretical quantization of gravity. In fact, these ideas can be recovered from the above formalism by regarding the central charge $c$ as divergent. Indeed, in eq. \eqref{orstoverlap} if one takes $cN\to \infty$ the right-hand side would rapidly oscillate, leading to zero.

The central charge diverges whenever the cut gives rise to a smooth classical geometry, $N\to\infty$. In the next section, we will offer a notion of quantum geometry, in which a cut, thought of as a representation of the diffeomorphism group, is interpreted as a finite dimensional representation. From this perspective then, the resolution of the problem of time is a truly quantum effect.

\section{Mesoscopic Quantum Gravity: the Embadon}\label{mol}

We have found that the central charge in the algebra of the operator $\hat C$ is infinite. We can track this divergence to the absence of a cutoff in the number of null generators, which is registered by the contribution $\delta^{(2)}(0)$ counting the number of points on the spatial cut. In this section, we suggest a model for how quantum gravity may evade this problem and render the central charge finite.

\subsection{Molecular Geometry}

At first sight, one might think that to render the central charge finite it is enough to introduce a cutoff in the theory. However, we are dealing with a theory of quantum gravity where, thanks to the equivalence principle, symmetries define the dynamics, and it is not acceptable to simply break the local symmetries by the presence of a cutoff. Instead, we need to change the nature of the vacuum to preserve the symmetries.

As shown in \cite{Donnelly:2016auv, Ciambelli:2021vnn}, the generic corner symmetry algebra induced from Einstein gravity at an isolated corner is $\text{ECS}:= \mathrm{Diff}(\cC)\ltimes SL(2,\mathbb{R})^\cC\ltimes (\RR^2)^{\cC}$. This is the starting point of the corner proposal \cite{Donnelly:2020xgu,  Donnelly:2022kfs,Freidel:2021cjp,Ciambelli:2021vnn,Ciambelli:2022cfr,Freidel:2015gpa, Geiller:2017xad, Speranza:2017gxd, Geiller:2017whh, Balasubramanian:2023dpj, Ciambelli:2024qgi} (see also the reviews \cite{Ciambelli:2022vot, Freidel:2023bnj, Ciambelli:2023bmn} and references therein). Here, as in \cite{Ciambelli:2023mir}, we are restricting our attention to a null hypersurface, and thus reducing this group to the so-called Weyl-BMS group $\text{BMSW}:= \mathrm{Diff}(\cC)\ltimes \mathbb{R}^{\cC}\ltimes \mathbb{R}^{\cC}$, where the first $\RR^\cC$ factor represents local boosts while the second one are the supertranslations. We expect that the quantum theory forms a representation of this symmetry group \cite{Ciambelli:2022vot, Freidel:2023bnj} or a deformation of it \cite{Donnelly:2022kfs}. We cannot simply break this symmetry with a cutoff since this would amount to breaking the underlying background independence. Quite remarkably the BMSW corner symmetry group (and related subgroups) is the group of symmetries of asymptotic infinity \cite{Campiglia:2014yka, Campiglia:2015yka, Compere:2018ylh, Freidel:2021fxf}.  More importantly for our interpretation, restricting our attention to the affine group $\text{GBMS}=\mathrm{Diff}(\cC)\ltimes \mathbb{R}^{\cC}$,  one remarks that this is the group of symmetries of a barotropic fluid \cite{holm2002euler,holm2001eulerpoincare, khesin2021geometric}, for which much is known about its representation \cite{Donnelly:2020xgu,Freidel:2024jyf}.\footnote{See \cite{Freidel:2024jyf} for a study of the asymptotic GBMS representation in terms of the barotropic fluid perspective and \cite{Barnich:2014kra, Barnich:2015uva} for analogous statements in 3d gravity. To include supertranslations, one needs to extend the phase space introducing edge modes on the cut. We plan to explore quantum representations of BMSW in future works, but it is clear that it will not change the area quantization proposal of this section.}

While we do not need to enter into the details of the representation theory of the hydrodynamic group GBMS, we can use it as a source of ideas for the present discussion. Indeed, we recall that the area density plays the role of a Casimir of GBMS and that there are essentially two kinds of representations depending on its properties. The first kind of representations are the classical fluid ones:  they are characterized by the fact that $\Omega$ is a strictly positive and absolutely continuous measure on the cut. 
This is the hypothesis that we implicitly have been making so far since $\Omega$ represents the induced area form, which is continuous and strictly positive by design in a classical spacetime.

 The second type of representations are the \emph{molecular representations}. The molecular representations are fundamentally quantum; they postulate that the area form operator $\hat{\Omega}(x)$ represents, when diagonalized, a positive but not strictly positive \emph{discrete measure} on the cut. The support of the measure represents the gravitational fluid constituents and determines how the boost symmetry $\mathbb{R}^{\cC}$ acts. In quantum hydrodynamics, these constituents are the fluid's molecules which, in addition, may carry quantum vortices labelling a $\mathrm{Diff}(\cC)$ representation \cite{Landau:1975pou}. 
 
Inspired by these results, this section aims to show that the \emph{molecular representation} of the area operator lead to a finite central charge. The terminology here reflects that one should think of the molecular representation as a description lying in between the Effective Field Theory (EFT) description, where geometry is classical, and a complete description of quantum gravity, such as string theory. The molecular description is a `mesoscopic' description that captures the essence of what it means for geometry to become quantum, without assuming that we have the full UV complete description of the theory, similar in spirit to the discovery of Brownian motion.

In the simplest molecular representation, one assumes that the   constituents are scalars. Concretely this means that we assume that $\hat\Omega$, as a quantum operator, has discrete support on the codimension-2 spatial cut, and is given by\footnote{We could have a more general description where the constituents are dipoles (or even more generally multipoles) in which case we would have higher multipole expansion $\Omega_i^{A} \pa_A\delta^{(2)}(z-z_i) + \Omega_i^{AB} D_AD_B\delta^{(2)}(z-z_i) +\ldots $ as constituents. We restrict our analysis to scalar constituents that represent area elements.  } 
\beq \label{molecular}
\hat\Omega(x)=\sum_{i=1}^N \hat\Omega_i(v) \delta^{(2)}(z-z_i),
\eeq
where $z_i$ are N distinct points on $\cC$, and $z_i\neq z_j$ when $i\neq j$.
The operator $\hat\Omega_i$ denotes the value of the area operator at the point $z=z_i$. Since the area commutes with itself, we can always work in a polarization where $\hat\Omega_i$ are diagonalized and treated as functions on the cut. Each puncture $z_i$ hosts a codimension-$2$ degree of freedom that carries quanta of area represented by $\hat\Omega_i$. We refer to these quanta of area as \emph{embadons}. The name embadon is a Greek term for area = \emb. The root of the word also reminds us of the word embedding, an accurate name for the geometry constituents. 

The first point to emphasize is that we have not introduced a UV regulator in terms of a fundamental length scale. We have introduced a dimensionless number $N$ that represents the number of fundamental constituents. We take this as a fundamental lesson of holography: the  regularization determines the number of constituents, not  a scale. What this buys us is the fact that this construction does not break the Diff$(\cC)$ gravitational symmetry. Indeed a diffeomorphism $\phi:\cC\to \cC$ on $\hat{\Omega}\to \phi^*\hat\Omega$   simply acts on the labels $z_i \to \phi^{-1}(z_i).$  A specific subclass of  diffeomorphism would simply permute the label points, i.e., $\phi^{-1}(z_i) = z_{\sigma(i)}$ where $\sigma \in S_N$ is a permutation. If we assume that the statistics of this field is bosonic, we can identify the configurations related by permutations. 

To reiterate, the fact that the molecular representation is compatible with diffeomorphism symmetry is what distinguishes it fundamentally from any other form of regularization, which would simply break the symmetry instead of representing it in a different manner than the continuum. In particular, the molecular representation is not a lattice discretization, nor is it a UV cutoff regularization. Instead, as the name suggests, it simply predicts the appearance of a constituent of the area, the embadon.

To understand the consequences of assuming a molecular  spectrum \eqref{molecular} for the area element, we analyze the symplectic structure. The spin-$0$ part of \eqref{ocan2} localizes entirely on the support of $\Omega$:\footnote{We here assume that $z_i$ are not part of the phase space, $\delta z_i=0$. This restriction can be relaxed including the Damour constraint \cite{Ciambellitoappear}.}
\beq\label{molOm0}
\Omega_{(0)}^{\mathsf{can}} =-\frac{1}{8\pi G}\int_{\cN}\ve_{{\cN}}^{(0)} \delta\mu \wedge \delta\Omega=-\frac{1}{8\pi G}\sum_{i=1}^N \int_{-\infty}^\infty \delta\mu_i(v) \wedge \delta\Omega_i(v)\rd v,
\eeq
where we denote $\mu_i(v):=\mu(v,z_i,\bar{z}_i)$.
The first piece of information we can extract from this presymplectic structure is that the values of $\mu(v,z,\bz)$ for $z\neq z_i$ are {\it pure gauge}. Only the value of $\mu$ at the location of the `punctures' $z_i$ are physical. Indeed, when given a symplectic form $\Omega^{\mathsf{can}}$  the gauge transformations are  the kernel of $\delta_X$: $I_{\delta_X} \Omega^{\mathsf{can}}=0$.

We are used to the classical condition that the area form is strictly positive. In this case the only gauge invariance available for the full symplectic form are diffeomorphism transformations. This is the usual notion of gauge invariance which is field independent.
When $\Omega$ is not strictly positive a new type of gauge invariance appears: any field transformation with value in the interior of the set $\{\Omega =0 \}$ is a gauge transformation. Hence the space of (gauge invariant) physical fields depends on the support of the area form $\Omega$. In other words the number of physical  degrees of freedom can be reduced by having a smaller support for the non-zero value of the area form.
The mechanism just described cannot happen in QFT since there, the geometry is a fixed background structure. 
It can only happen in quantum gravity where the area element is a physical field and the question of what support it has depends on the quantum gravitational state. Here we investigate the `molecular' case where the support of $\Omega$ is along a finite number of points on the cut. In this case the set of spin-$0$ physical degrees of freedom is \emph{finite dimensional}.

From \eqref{molOm0} we  easily conclude that the  Poisson bracket of the spin-$0$ data is 
\beq
\{\Omega_i(v_1),\mu_j(v_2)\}=8\pi G \,\delta(v_{12})\delta_{ij}.
\eeq 
Thanks to ultralocality, the exact same procedure as in \eqref{mphi} can be implemented here for the quantization of each molecular component, leading to the OPE
\beq
\hat\mu_i(v_1)\hat\Omega_j(v_2)\sim -\frac{4G\hbar \delta_{ij}}{v_{12}-i\epsilon}.
\eeq
The molecular representation of symplectic degrees of freedom extends to the spin-$2$ sector in \eqref{ocan2}
\beq\label{can2}
\Omega^{\mathsf{can}}_{(2)} 
= \frac{1}{16 \pi G}\sum_{i=1}^N \int_{-\infty}^\infty  \rd v\Big(\delta\left(\Omega_i \sigma_i^{ab}\right)\wedge\delta q_{iab}\Big),
\eeq
where $\sigma^{ab}_i := \sigma^{ab}(v,z_i,\bar{z}_i)$ and $q_{iab}(v) := q_{ab}(v,z_i,\bar{z}_i)$. From this we conclude that the value of $q_{ab}$ and its time derivative outside of the punctures are pure gauge.
Similar conclusions can be achieved for the matter sector.

The generator of time reparametrizations can be constructed in this restricted phase space. Since we have established that the field variations outside the punctures are pure gauge, the only relevant action of the diffeomorphisms $\xi=f\pa_v$ on the molecular phase space is given by the action of its restriction $f_i(v)=f(v,z_i,\bz_i)$ on each molecular variable.
The canonical primed action is the molecular restriction of (\ref{trs}-\ref{sitr}),
\beq\label{trsmol}
&\fL_{\hat{f}}'q_{iab}=\cL_{f_i} q_{ 
      iab},\qquad
\fL_{\hat{f}}' \sigma_{ia}{}^b =   \pa_v( f_i\sigma_{ia}{}^b),&\\
&\fL'_{\hat{f}}\mu_i=\pa_v(f_i\mu_i)+\pa_v^2 f_i,\qquad
\fL'_{\hat{f}}\Omega_i=f_i\pa_v\Omega_i,\qquad
\fL'_{\hat{f}}\theta=\pa_v(f_i\theta),&
\eeq
and one finds that its canonical generator is simply given by
\be 
C_i(v)=\pa_v^2\Omega_i-\mu_i\pa_v\Omega_i+\Omega_i(\sigma_i^{ab}\sigma_{iab}+8\pi G T^{\mathrm{mat}}_{ivv}).
\ee 
Ultralocality plays a crucial role, resulting in the constraint localizing at punctures, $\{ C_i , C_j \}\propto \delta_{ij}$. 

At the quantum level, using
\beq
\hat C_i(v)=\frac{C_i(v)}{4G\hbar},
\eeq
the non-perturbative OPE is of the form
\be 
\hat C_i(v_1) \hat C_j(v_2)\sim  \delta_{ij}\left( 
\frac{c}{2(v_{12}-i\epsilon)^4}+\frac{2 \hat C_j(v_2)}{(v_{12}-i\epsilon)^2}
+\frac{\pa_{v_2} \hat C_j (v_2)}{v_{12}-i\epsilon}
\right),
\ee 
where $c$ is the single null ray central charge. Notice that, as stated, the construction is diffeomorphism covariant on the base, as $\phi:\cC\to \cC$ simply acts on the labels $z_i \to \phi^{-1}(z_i)$. However, to realize on the phase space this symmetry one has to include the Damour constraint, which is the generator of such diffeomorphisms. This will be discussed in our upcoming paper \cite{Ciambellitoappear}, but the underlying assumption is that embadons will carry through. The embadons which are localized on each cut of $\cN$ describe the number of constituents associated with this cut. Therefore, what we have achieved with the introduction of embadons is that the total central charge of the cut, associated with the constraint $\hat{C}=\sum_{i=1}^N \hat{C}_i$ is now finite and given by $c N$, where $N$ is the total number of embadons.

In conclusion, we have studied a regime called mesoscopic quantum gravity where we encountered the appearance of area constituents, called embadons. Embadons allow us to localize the symplectic analysis and quantization on punctures, making the total central charge finite.

\subsection{Relationship with Other Approaches to Quantum Gravity}

In this section, we want to clarify the connections between the ideas developed here and previous results in the quantum gravity literature. Let us first emphasize that our approach in this manuscript is a `bottom-up' approach to quantum gravity. This means that we are proposing an \emph{effective} description that includes backreaction of matter and spin-$2$ degrees of freedom on geometry. In this regard, it goes beyond the usual QFT and classical gravity perturbative regime, which assumes that geometry is fixed. 
In our description, embadons are representation labels of a fluid-like symmetry group. The area operator is expected to be related to the modular Hamiltonian, which, in turn, measures the entanglement of states. Consequently, since the corner symmetry group
organizes embadon representations, it is connected to the entanglement structure of the theory. This `mesoscopic' description of quantum geometry is not a fundamental description of nature like string theory aims to be. It is a new stepping stone towards full quantum gravity that better equips us to deal with the upcoming technical, conceptual, and experimental challenges. 

The starting point of UV-complete quantum gravity theories, which one could call `top-down' models, is often to postulate desirable features of gravity in the UV and then try to connect back to QFT and general relativity in the IR. In one way or another, these theories must take into account the classical symmetries of the phase space. Our bottom-up approach has the virtue of taking these classical symmetries as the starting point, proposing a covariant quantization. The final aim is to appreciate the presence of a regime, the mesoscopic regime, where one can consistently straddle the gap between top-down and bottom-up models. Our approach integrates two essential lessons coming from different quantum gravity communities. The first lesson is from loop quantum gravity (LQG) \cite{Ashtekar:1986yd, Rovelli:1995ac, Thiemann:2007pyv, Perez:2012wv}, and is that \emph{geometry is quantum}. The second lesson comes from string theory and AdS/CFT \cite{Maldacena:1997re, Witten:1998qj, Gubser:1998bc, Aharony:1999ti}, and is that \emph{gravity is holographic}. These two approaches illuminate different aspects of quantum gravity, and we propose that a comprehensive understanding of quantum gravity can be achieved by integrating them coherently.

First, in LQG, the states are assumed to be supported by graphs called spin networks whose edges carry quanta of area \cite{Rovelli:1994ge, Ashtekar:1996eg, Ashtekar:2004eh}. It was then proposed that the counting of surface states associated with the intersection of spin networks with $2$-horizons gives rise to the black-hole entropy  \cite{Smolin:1995vq,Krasnov:1996tb,Krasnov:1996wc,Ashtekar:2000eq, Domagala:2004jt, Engle:2009vc, Ghosh:2011fc, Ghosh:2014rra}. This leads to a picture of quantum geometry for black holes that is consistent with the more general embadon picture proposed here.

In AdS/CFT, and more broadly in holography, the fundamental degrees of freedom are, by definition, excitations of a boundary CFT. Therefore, they are, like embadons, localized on codimension-$2$ slices. This is the outcome of the holographic dictionary: bulk asymptotic symmetries are dual to boundary global symmetries. Consequently, physical charges are codimension-$1$ integrals on the boundary and, therefore, codimension-$2$ extended quasilocal charges in the bulk. Moreover, the entanglement of boundary quantum states builds the architecture of bulk spacetime, and its smooth gluing of subregions \cite{VanRaamsdonk:2010pw, Faulkner:2013ana, Almheiri:2014lwa, Jafferis:2015del, Dong:2016hjy, Pastawski:2015qua, May:2022clu}. In particular, this is based on the correspondence where the minimal area in the bulk is dual to the boundary entanglement entropy \cite{Ryu:2006bv}, thereby highlighting the special role of bulk areas and entanglement wedges.

In LQG, there are three significant areas for potential improvement, which incidentally are the lessons of holography. The first one is more technical: apart from works on black-hole entropy, the focus of LQG has been on promoting the fundamental degrees of freedom to be codimension-$1$ extended excitations rather than codimension-$2$. The second is that in LQG, quantum geometry excitations carry a representation of the SU$(2)$ local Lorentz symmetry only, which is assumed to be the only local group responsible for the entanglement of excitations, while no entanglement arises from diffeomorphisms. Nonetheless, it is now firmly established \cite{Donnelly:2016auv, Freidel:2020xyx} that diffeomorphisms must carry entanglement in quantum gravity. 
This follows from the fact that the gluing of adjacent regions along codimension-$2$ spacetime cuts is achieved at the quantum level as a fusion product that ensures the matching of Noether charges associated with each symmetry, including diffeomorphisms tangent or transverse to the cut. The third and most significant opportunity for improvement in LQG is related to this issue: the loop gravity vacuum representing empty spacetime makes no reference to a choice of holomorphic polarization and is consequently not reconcilable with a Fock ground state, \cite{Thiemann:2004qu,Thiemann:2024uwr}. In particular, the modular Hamiltonian associated with cuts of quantum geometry in LQG is not related to the boost diffeomorphism symmetry (see \cite{Ghosh:2011fc, Bianchi:2012ui} for some notable exceptions), which is one of the foundational principles of QFT \cite{Bisognano:1976za, Unruh:1976db, Casini:2010kt, Casini:2022rlv}, and in particular of holography \cite{Casini:2011kv, Blanco:2013joa, Jafferis:2015del, Faulkner:2016mzt}.

In holography, there are significant areas for potential improvement, which incidentally are the lessons of LQG. Indeed, in most holographic treatments, the area is regarded as a central element and often not as an operator. This is in line with the idea in holography that gravity can be quantized without quantizing geometry. While this is a foundational principle of the AdS/CFT correspondence, it is also a point of contention and a potential area for improvement for several reasons. First, while AdS/CFT provides a robust framework, it does not fully resolve the issue of how to reconcile quantum principles with the fundamental nature of spacetime geometry at a microscopic level. Quantizing gravity involves dealing with the quantum fluctuations of spacetime itself or with the quantum backreaction on the geometry of field perturbation, which is not fully addressed in the holographic duality. Second, gravity in the bulk AdS space emerges from the boundary CFT, suggesting that geometry may not be fundamental but rather an effective description. This raises questions about the underlying nature of space and time at the quantum level and whether gravity truly can be understood without a deeper quantization of geometry. Related to this is the proposition that the holographic nature of gravity is a quasilocal feature \cite{Donnelly:2016auv,Ciambelli:2021nmv, Balasubramanian:2023dpj,Klinger:2023auu} that can be extended to finite regions using corner symmetries. The AdS/CFT correspondence focuses on observables  defined only at infinity (with few exceptions \cite{Bousso:2015mna}), and thus understanding the nature of quantities like entanglement entropy or information in terms of bulk geometry remains a challenge \cite{May:2022rko}. Clarifying how these dual descriptions relate more intuitively to traditional semiclassical gravitational concepts could provide deeper insights into the nature of gravity itself.

As highlighted, a notable feature is that one perspective contributes to improving the deficiencies of the other. First, both perspectives lead to the conclusion that gravity possesses codimension $2$ excitations and that the black hole entropy is finite; hence, the fundamental excitations should regulate UV divergences. From LQG, we learn that these excitations should carry quanta of area, necessitating the promotion of the area to a quantum operator. In parallel, holography makes concrete the connection between the area operator, modular Hamiltonian, and entanglement entropy. These elements naturally interconnect when considering the quantization of null geometry. Specifically, the emergence of embadons at the mesoscopic level integrates the diverse insights from top-down quantum gravity theories comprehensively.

\subsection{Embadons and Gravitons}

We have seen in Section \ref{sec: asymptinfty} that the dynamics and operator structure of a large null surface becomes, in asymptotically flat space, the same as the Bondi description of asymptotic infinity. The shear becomes the news $\sigma_{AB}\to N_{AB}/r$ while, looking at constant cuts $u=T$, the renormalized area $\Omega_R =\Omega-\frac{(u-u_0)^2}{4}$ becomes the  time integral of the Bondi mass. For these cuts, where $D_AT=0$,  we have from \eqref{or} that 
\be \label{aren2}
\Omega_R(T)= - 2 \int_{T}^{+\infty} (u-T) T^{\mathsf{H}} + \frac12 \int_{T}^{+\infty} (u-T) D_AD_B N^{AB}.
\ee
As already remarked, this formula shows that the integral of $\Omega_R$ on the cut is strictly negative if the Bondi mass is non-vanishing. Flat space, whereby $m_B=0$, is thus a \emph{maximal area} spacetime. Any other physical asymptotic states representing a non-empty spacetime have a lower asymptotic area. Stated differently, the flat space background is realized as a state of  \emph{maximal} embadon density. 

The asymptotic limit therefore is such that $N\to \infty$, where $N$ controls the finite central charge. The difference between $N\to \infty$ at a finite distance hypersurface and asymptotic null infinity is that in the latter we also have that the area $A = \sum_{i=1}^N \Omega_{i}$ diverges. Conversely, the QFT limit for bulk hypersurfaces is set by $N\to \infty$ while keeping $A$ finite. This suggests that in the QFT limit $\Omega_i \propto \rho/N$, where $\rho$ becomes the continuum area element.

We thus understand the asymptotic quantum gravity vacuum as a continuum limit of embadons, in the same way that classical fluid dynamics can be understood as a thermodynamic limit of molecules. What is interesting is that in this limit, the flat space vacuum acts as a maximal density state, similar to a Fermi surface at maximal density. In this limit, all the excitations lower the area element eigenvalues, as in a graviton emission  process. More precisely, we know \cite{Raclariu:2021zjz, Pasterski:2021rjz} that the Bondi news can be expanded in terms of graviton creation and annihilation operators $(a_\pm(\omega), a^\dagger_\pm(\omega)) $, while since $T^{\mathsf{H}}(u)$ is quadratic in the news it can be expanded as a quadratic functional in $(a_\pm(\omega), a^\dagger_\pm(\omega))$. The formula \eqref{aren2} then provides an explicit action of the area element acting on the Fock ground state. The detailed study of this action is beyond the scope of the present paper. What we can conclude, nevertheless, is that while graviton eigenstates are not area eigenstates, coherent graviton states are associated with area element coherent states. Therefore, we see a direct connection between embadon coherent states and those of gravitons.

While the limit to asymptotic null infinity makes the  connection between embadons and gravitons sharp, we remark that such a connection is also present  for a finite-distance null hypersurface in the bulk. This is visible both in the classical phase space and in the dynamics. In the former, the evidence in support of this claim is the bracket above \eqref{omOm}, which shows that the fluctuation of the perturbative area is dictated by the perturbative gravitons $\pa_v X$ and $\pa_v\bX$.
 In the latter, this is readily deduced from the Raychaudhuri constraint \eqref{RC1}. Suppose for instance that we are in dressing time $\mu=0$ and there is no matter present,  the  area is  the ANEC operator $\int_{-\infty}^\infty \rd v\, v T$ (see \eqref{Tfirst}), which is the light-ray operator describing the gravitons on the null hypersurface.

\section{Final Words}\label{finale}

We have explored the quantization of gravity on a null hypersurface. We have shown that the ultralocal nature of the phase space, combined with the null geometric structure, reveals that gravity on a null hypersurface effectively reduces to a one-dimensional chiral sector of a CFT. This insight allowed us to predict the presence of a central charge and, consequently, an anomaly in the quantum primed time reparametrization due to the rescaling of the null generator needed to establish it as a background structure. We computed the central charge and demonstrated that it quantifies the degrees of freedom on each null generator, formally diverging when infinitely many such generators are considered. We studied this concept in various setups: the spin-0 sector, the perturbative gravity regime on static and expanding backgrounds, and asymptotic infinity. 

We then noted that the presence of the central charge allows for a resolution of the problem of time in quantum gravity. This led us to  venture into a novel and exploratory direction: we proposed a mesoscopic regime where `molecules' of geometry arise. In this regime, the quantum area operator acquires a discrete spectrum, with constituents that we called embadons, from the Greek word \emb = area. Notably, we showed that in this mesoscopic regime, the central charge can be rendered finite, which we suggest should be taken to be the hallmark of quantum gravity, outside of semiclassical constructions. 

There are many avenues our exploration has touched upon that are worth further development. 
First, in a recent series of papers \cite{Verlinde:2019ade, Verlinde:2022hhs} Verlinde and Zurek  have proposed that in quantum gravity, fluctuations of the modular Hamiltonian $K$ are tamed in a universal manner. Instead of being infinite, it satisfies $(\Delta K)^2 = \langle K\rangle$. An experiment designed to measure this effect has been proposed \cite{Vermeulen:2024vgl}. It is worth mentioning that the property of the modular Hamiltonian cited above is a well-known property of 2d CFTs (this was mentioned recently, for example, in \cite{Banks:2024cqo} and will be explored in \cite{robtoappear}). In this paper, we have demonstrated that the dynamics of quantum gravity on a null hypersurface is controlled by a chiral CFT, and so our construction suggests that there is good reason to believe that  $(\Delta K)^2/\langle K\rangle$ is fixed in quantum gravity in general. The mesoscopic construction is a framework in which both quantities are expected to be finite, as a consequence of the central charge being finite. As we discussed, the finiteness of the central charge allows us to consider superposition of states in the vacuum module. Conversely, this is not possible in the QFT limit, where the central charge diverges and thus one is forced to have superselection. As such, this is a truly a quantum gravity effect, that we intend to explore in future work. 

The perturbative setup on expanding backgrounds is definitely worth pursuing, in particular its canonical quantization. It would be interesting to complete the deformation quantization analysis, and compute explicitly the star product of constraints. Mastering these backgrounds allows us to make contact with the thriving topic of de Sitter perturbation theory and cosmology. 

While we worked primarily in $4$-dimensional bulks, our results hold in any dimension. We could for instance explore $2$-dimensional gravity, where the cut becomes a point, and there is no need to appeal to ultralocality. Another interesting venture is to apply our results to the dimensional reduction of gravity in the vicinity of a null hypersurface, such as a black hole horizon. This allows us to make contact with the thermodynamic analysis of black holes, and extend its domain of applicability to all
null hypersurfaces. We expect that our construction enhances the thermodynamic laws to hold in the regime where quantum geometric effects are relevant. For black hole horizons, our results should align with \cite{Solodukhin:1998tc, Carlip:2017eud, Almheiri:2019yqk, Adami:2021kvx}, allowing to export their analysis to any  null hypersurface, and making this construction independent of the symmetry structure of the background chosen. Furthermore, we intend to relate in detail our work with the near-horizon CFT construction in  \cite{Carlip:1998wz, Solodukhin:1998tc}, see also \cite{Banks:2021jwj}.

To describe all the constraints induced by gravity on a $3$-dimensional null hypersurface, we must include the Damour constraint into the quantization picture described here. This is a priority for us, and we expect to report about it soon. Incidentally, the comprehension of the Damour constraint makes further contact with the celestial CFT program, where the Virasoro algebra arises in the spatial directions on the celestial sphere \cite{Kapec:2016jld, Strominger:2017zoo}. Including the Damour constraint is expected to have repercussions for the embadon, providing a way to formulate the embadon representation theory on the cut, and its interplay with the Raychaudhuri constraint.

The discussion in Section \ref{qutime} touched upon foundational properties of QFT. We wish to explore how our analysis intertwines with the Hadamard property of the vacuum \cite{Brunetti:1995rf,Radzikowski:1996pa,Brunetti:1999jn, Hollands:2001nf}, the  Hadamard singularity structure
\cite{fulling1978singularity, fulling1981singularity}, and its null limit \cite{kay1991theorems}. Another 
aspect that we wish to explore further is the rephrasing of the anomaly found here in the standard QFT formalism \cite{Alvarez-Gaume:1983ihn,Weinberg:1995mt, Bilal:2008qx}. In this direction, we could deepen our understanding of the phase space and its quantization arising from our symplectic potential, which has a universal structure appearing in many physics systems, such as non-linear sigma models and thus string theory.

A final direction to follow is the interplay between our quantization framework and quantum information tools in gravity. Specifically, we plan to study the generalized second law on a generic null hypersurface. The notion of time chosen becomes of primary importance in formulating the generalized second law, and subsequently the quantum focusing conjecture \cite{Bousso:2015mna}. Indeed, the latter is not invariant under rescaling, and this might provide yet another way to corroborate the special role played by the conformal time discussed in this paper.

\paragraph{Acknowledgement} We are thankful to Ivan Agullo, Abhay Ashtekar, Ana-Maria Raclariu, Gautam Satishchandran, Antony Speranza, Simone Speziale, Aron Wall, and Kathryn Zurek for  discussions and constructive criticisms.
LC thanks Miguel Campiglia and Juan Maldacena for important discussions on related projects. Similarly, RGL thanks Marc Klinger for discussions on ongoing related collaborations. We are grateful to BIRS (Banff) for the warm hospitality during our focused research group initiatives of November 2022, where this work was initiated, and of November 2023.
Research at Perimeter Institute is supported in part by the Government of Canada through the Department of Innovation, Science and Economic Development Canada and by the Province of Ontario through the Ministry of Colleges and Universities. The work of RGL is partially supported by the U.S. Department of Energy under contract DE-SC0015655, and RGL thanks the Perimeter Institute for supporting collaborative visits. This work was supported by the Simons Collaboration on Celestial Holography. 

\appendix
\renewcommand{\theequation}{\thesection.\arabic{equation}}
\setcounter{equation}{0}

\section{Classical Details}\label{class}

In this appendix, we collect results from \cite{Ciambelli:2023mir} that will be useful at various points in the main text. If we assume no fluxes at $\partial \cN$, we can derive the kinematic Poisson bracket from the symplectic two form \eqref{ocan2}. To do so, we use the Beltrami differentials for the unimodular metric $\bq_{ab}$. The Beltrami diffentials are the fields $\zeta(x)$ and $\bzeta(x)$ satisfying
\beq\label{bp}
q_{ab}=\Omega \bq_{ab}, \qquad \bq_{ab}\rd x^a \rd x^b=2\frac{\vert \rd z+\zeta \rd \bz\vert^2}{\beta},
\eeq
where $\beta=1-\zeta\bzeta$. They allow us to rewrite the spin-$2$ degrees of freedom in terms of $\zeta, \bzeta$, and their temporal derivatives\footnote{Note the useful identity $\sigma_a{}^b\sigma_b{}^a=2\Delta_v\bD_v$.}
\beqn
\Delta_v=\frac{\pa_v\zeta}{\beta},\qquad \bD_v=\frac{\pa_v\bzeta}{\beta},\qquad \omega_v=\frac{\zeta\pa_v\bzeta-\bzeta\pa_v\zeta}{2\beta}=-\overline{\omega}_v.
\eeqn

Then, introducing the propagator ($\cP_{12}=-\bcP_{21}$)
\beq\label{propa}
\bcP_{12}=\frac{e^{-2\int_{v_1}^{v_2}\omega_v \rd v}}{\sqrt{\Omega_1\Omega_2}}
H(v_1-v_2)\delta^{(2)}(z_1-z_2),
\eeq
with $H(v_{12})$ the odd Heaviside function satisfying $\pa_{v_1} H(v_{12}) =\delta(v_{12})$,
we derived in \cite{Ciambelli:2023mir} the kinematic Poisson brackets
\beqn
\{\Omega_1,\mu_2\}&=&8\pi G\,\delta^{(3)}(x_{12})
\label{omu}
\\
\label{mumu}
\{\mu_1,\mu_2\}&=&4\pi G\left(\Delta_{v_1}\bD_{v_2}\bcP_{12}+\bD_{v_1}\Delta_{v_2}\cP_{12}\right)
\\
\{\mu_1,\zeta_2\}&=&-4\pi G\,\Delta_{v_1} \beta_2\bcP_{12}
\\
\{\zeta_1,\bzeta_2\}&=&4\pi G\,\beta_1\beta_2\cP_{12},
\label{lastp}
\eeqn
where we use the notation $\Omega_1= \Omega(x_1)$, etc.

\section{Perturbative Brackets}\label{A2}

To find the Poisson bracket, it is convenient to use the composite functional $\Omega_B^\mu$ in the phase space basis, such that the latter is spanned by $Z^\alpha(x)=(\omega(x),\Omega_B^\mu(x),X(x),\bX(x))$. Then, we can write the symplectic two form \eqref{s2p} as a bi-local integral
\beq
\Omega^{\mathrm{can}}=\frac12\int_{\cN_1}\ve_{\cN_1}^{(0)}\int_{\cN_2}\ve_{\cN_2}^{(0)}\delta Z^\alpha(x_1)\wedge \Omega_{\alpha\beta}^{\mathrm{can}}(x_1,x_2)\delta Z^\beta(x_2),
\eeq
with ($\delta_{12}:=\delta^{(3)}(x_{12})$)
\beq
\Omega_{\alpha\beta}^{\mathrm{can}}(x_1,x_2)=\begin{pmatrix}
0 & \pa_{v_2}\left(\frac{\pa_{v_2}\delta_{12}}{\pa_{v_2}\Omega^\mu_{B2}}\right) & 0 & 0\\
-\pa_{v_1}\left(\frac{\pa_{v_1}\delta_{12}}{\pa_{v_1}\Omega^\mu_{B1}}\right) & 0 & \pa_{v_2}\bX_2 \delta_{12} & \pa_{v_2}X_2 \delta_{12}\\
0 & -\pa_{v_1}\bX_1 \delta_{12} & 0 & -2\sqrt{\Omega^\mu_{B1}\Omega^\mu_{B2}}\pa_{v_1}\delta_{12}\\
0 & -\pa_{v_1}X_1 \delta_{12} & 2\sqrt{\Omega^\mu_{B1}\Omega^\mu_{B2}}\pa_{v_2}\delta_{12} & 0
\end{pmatrix}.
\eeq
To check this, one uses $\mu=\pa_v\ln \pa_v\Omega_B^\mu$ and thus $\delta\mu=\pa_v\frac{\pa_v\delta\Omega_B^\mu}{\pa_v\Omega_B^\mu}$. 

We then invert this matrix, and obtain
\beq
\Omega^{\alpha\beta}_{\mathrm{can}}(x_1,x_2)=\begin{pmatrix}
\Omega^{\omega \omega}_{\mathrm{can}}(x_1,x_2) & -\Delta\Omega^\mu_{B 21}\Theta(v_{21}) & \Omega^{\omega X}_{\mathrm{can}}(x_1,x_2) & \Omega^{\omega\bX}_{\mathrm{can}}(x_1,x_2) \\
\Delta\Omega^\mu_{B 21}\Theta(v_{21}) & 0 & 0 & 0\\
-\Omega^{\omega X}_{\mathrm{can}}(x_1,x_2) & 0 & 0 & \frac{\Theta(v_{21})}{2\sqrt{\Omega_{B1}^\mu\Omega_{B2}^\mu}}\\
-\Omega^{\omega\bX}_{\mathrm{can}}(x_1,x_2) & 0 & \frac{\Theta(v_{21})}{2\sqrt{\Omega_{B1}^\mu\Omega_{B2}^\mu}} & 0
\end{pmatrix}\delta^{(2)}(z_{12}),
\eeq
with $\Delta\Omega^\mu_{B 21}=\Omega^\mu_{B}(x_2)-\Omega^\mu_B(x_1)$, and\footnote{We used $\Delta\Omega^\mu_{B 31}$ for compactness, but the spatial points are the same here: $\Omega^\mu_B(v_3,z_1,\bz_1)-\Omega^\mu_B(v_1,z_1,\bz_1)$.}
\beqn
&\Omega^{\omega\bX}_{\mathrm{can}}(x_1,x_2)=\frac{\Theta(v_{21})}{\sqrt{\Omega^\mu_{B}(x_2)}}\int_{v_1}^{v_2}\frac{\pa_{v_3}\bX(v_3,z_1,\bz_1)}{2\sqrt{\Omega^\mu_{B}(v_3,z_1,\bz_1)}}\Delta\Omega^\mu_{B31}\rd v_3&\\
&\Omega^{\omega X}_{\mathrm{can}}(x_1,x_2)=\frac{\Theta(v_{21})}{\sqrt{\Omega^\mu_{B}(x_2)}}\int_{v_1}^{v_2}\frac{\pa_{v_3}X(v_3,z_1,\bz_1)}{2\sqrt{\Omega^\mu_{B}(v_3,z_1,\bz_1)}}\Delta\Omega^\mu_{B31}\rd v_3&\\
&\Omega^{\omega \omega}_{\mathrm{can}}(x_1,x_2)=\Theta(v_{21})\int_{v_1}^{v_2}\pa_{v_3}\Omega^\mu_{B3}\rd v_3\int_{v_1}^{v_3}\rd v_4\int_{v_1}^{v_4}\rd v_5\frac{\Delta\Omega^\mu_{B51}}{2\sqrt{\Omega^\mu_{B4}\Omega^\mu_{B5}}}\left(\pa_{v_5}\bX_5\pa_{v_4}X_4+\pa_{v_5}X_5\pa_{v_4}\bX_4\right)\nonumber&
\eeqn
Furthermore, we introduced $\Theta(v_{12})$ satisfying 
\beq\label{thhe}
v_1>v_2 \quad \Rightarrow \quad \Theta(v_{12})=1,\qquad v_1<v_2 \quad \Rightarrow \quad \Theta(v_{12})=0.
\eeq

Using the conventions $\Omega^{\alpha\beta}_{\mathrm{can}}(x_1,x_2)=-\{Z^{\alpha}(x_1),Z^\beta(x_2)\}$, we read the Poisson bracket
\beqn
\{\omega_1,\omega_2\}=-\int_{v_1}^{v_2}\rd v_3\int_{v_1}^{v_3}\rd v_4\int_{v_1}^{v_4}\rd v_5\frac{\pa_{v_3}\Omega^\mu_{B3}\Delta\Omega^\mu_{B51}}{2\sqrt{\Omega^\mu_{B4}\Omega^\mu_{B5}}}\left(\pa_{v_5}\bX_5\pa_{v_4}X_4+\pa_{v_5}X_5\pa_{v_4}\bX_4\right)\Theta(v_{21})\delta^{(2)}(z_{12}),\nonumber
\eeqn
together with
\beqn
\{\omega_1,\Omega_{B2}^\mu\}&=&\Delta\Omega^\mu_{B 21}\Theta(v_{21})\delta^{(2)}(z_{12})\\
\{\omega_1,X_2\}&=&-\int_{v_1}^{v_2}\frac{\pa_{v_3}X_3}{2\sqrt{\Omega^\mu_{B2}\Omega^\mu_{B3}}}\Delta\Omega^\mu_{B31}\rd v_3 \ \Theta(v_{21})\delta^{(2)}(z_{12})\\
\{\omega_1,\bX_2\}&=&-\int_{v_1}^{v_2}\frac{\pa_{v_3}\bX_3}{2\sqrt{\Omega^\mu_{B2}\Omega^\mu_{B3}}}\Delta\Omega^\mu_{B31}\rd v_3 \ \Theta(v_{21})\delta^{(2)}(z_{12})\\
\{X_1,\bX_2\}&=&-\frac{\Theta(v_{21})\delta^{(2)}(z_{12})}{2\sqrt{\Omega^\mu_{B1}\Omega^\mu_{B2}}}.
\eeqn
These are the brackets reported in the main body of the paper. Some useful composite brackets are
\beqn
\{\pa_{v_1}\omega_1,\Omega^\mu_{B2}\}&=&-\pa_{v_1}\Omega_{B1}^\mu\Theta(v_{21})\delta^{(2)}(z_{12}),\\
\{\pa^2_{v_1}\omega_1,\Omega^\mu_{B2}\}&=&-\left(\pa^2_{v_1}\Omega_{B1}^\mu\Theta(v_{21})-\pa_{v_1}\Omega_{B1}^\mu\delta(v_{21})\right)\delta^{(2)}(z_{12}),\\
\{\pa_{v_1}\omega_1,X_2\}&=&\pa_{v_1}\Omega^\mu_{B1}\int_{v_1}^{v_2}\frac{\pa_{v_3}X_3}{2\sqrt{\Omega^\mu_{B2}\Omega^\mu_{B3}}}\rd v_3\Theta(v_{21})\delta^{(2)}(z_{12})\\
\{\pa_{v_1}\omega_1,\pa_{v_2}X_2\}&=&\pa_{v_1}\Omega^\mu_{B1}\left(\frac{\pa_{v_2}X_2}{2\Omega^\mu_{B2}}-\frac{\pa_{v_2}\Omega^\mu_{B2}}{2\Omega^\mu_{B2}}\int_{v_1}^{v_2}\frac{\pa_{v_3}X_3}{2\sqrt{\Omega^\mu_{B2}\Omega^\mu_{B3}}}\rd v_3\right)\Theta(v_{21})\delta^{(2)}(z_{12})\\
\{\pa^2_{v_1}\omega_1,X_2\}&=&\left(\pa^2_{v_1}\Omega^\mu_{B1}\int_{v_1}^{v_2}\frac{\pa_{v_3}X_3}{2\sqrt{\Omega^\mu_{B2}\Omega^\mu_{B3}}}\rd v_3-\frac{\pa_{v_1}\Omega^\mu_{B1}\pa_{v_1}X_1}{2\sqrt{\Omega^\mu_{B2}\Omega^\mu_{B1}}}\right)\Theta(v_{21})\delta^{(2)}(z_{12})\\
\{\pa^2_{v_1}\omega_1,\pa_{v_2}X_2\}&=&\pa^2_{v_1}\Omega^\mu_{B1}\left(\frac{\pa_{v_2}X_2}{2\Omega^\mu_{B2}}-\frac{\pa_{v_2}\Omega^\mu_{B2}}{2\Omega^\mu_{B2}}\int_{v_1}^{v_2}\frac{\pa_{v_3}X_3}{2\sqrt{\Omega^\mu_{B2}\Omega^\mu_{B3}}}\rd v_3\right)\Theta(v_{21})\delta^{(2)}(z_{12})\nonumber\\
&&+\frac{\pa_{v_2}\Omega^\mu_{B2}\pa_{v_1}X_1}{2\Omega^\mu_{B2}}\left(-\delta(v_{12})+\frac{\pa_{v_1}\Omega^\mu_{B1}}{2\sqrt{\Omega^\mu_{B1}\Omega^\mu_{B2}}}\Theta(v_{21})\right)\delta^{(2)}(z_{12})
\eeqn
\beqn
\{\pa_{v_1}\omega_1,\pa_{v_2}\bX_2\}&=&\pa_{v_1}\Omega^\mu_{B1}\left(\frac{\pa_{v_2}\bX_2}{2\Omega^\mu_{B2}}-\frac{\pa_{v_2}\Omega^\mu_{B2}}{2\Omega^\mu_{B2}}\int_{v_1}^{v_2}\frac{\pa_{v_3}\bX_3}{2\sqrt{\Omega^\mu_{B2}\Omega^\mu_{B3}}}\rd v_3\right)\Theta(v_{21})\delta^{(2)}(z_{12})\\
\{\pa^2_{v_1}\omega_1,\bX_2\}&=&\left(\pa^2_{v_1}\Omega^\mu_{B1}\int_{v_1}^{v_2}\frac{\pa_{v_3}\bX_3}{2\sqrt{\Omega^\mu_{B2}\Omega^\mu_{B3}}}\rd v_3-\frac{\pa_{v_1}\Omega^\mu_{B1}\pa_{v_1}\bX_1}{2\sqrt{\Omega^\mu_{B2}\Omega^\mu_{B1}}}\right)\Theta(v_{21})\delta^{(2)}(z_{12})\\
\{\pa_{v_1}\omega_1,\pa_{v_2}\omega_2\}&=&-\pa_{v_1}\Omega_{B1}^\mu\pa_{v_2}\Omega_{B2}^\mu K(x_1,x_2)\Theta(v_{21})\delta^{(2)}(z_{12})
\eeqn
where we used $\pa_{v_1}\Theta(v_{21})=-\delta(v_{12})$, and introduced the spin-$2$ holonomy functional
\beq
K(x_1,x_2)=\int_{v_1}^{v_2}\rd v_3\int_{v_1}^{v_3}\rd v_4\frac{\pa_{v_4}\bX_4\pa_{v_3}X_3+\pa_{v_4}X_4\pa_{v_3}\bX_3}{2\sqrt{\Omega^\mu_{B3}\Omega^\mu_{B4}}}.
\eeq
This functional satisfies the useful identities $K(
x_1,x_1)=0=K(x_2,x_2)$, and
\beq
\pa_{v_1}K(x_1,x_2)=-\int_{v_1}^{v_2}\rd v_3 \frac{\pa_{v_1}X_1\pa_{v_3}\bX_3+\pa_{v_1}\bX_1\pa_{v_3}X_3}{2\sqrt{\Omega_{B3}^\mu\Omega_{B1}^\mu}}.
\eeq

\bibliographystyle{uiuchept}
\bibliography{ARXIVv2.bib}

\end{document}